\documentclass[preprint]{elsarticle}

\usepackage[colorlinks=true,linkcolor=black, citecolor=blue, urlcolor=blue]{hyperref}

\usepackage{color}
\usepackage{hyperref}
\usepackage{booktabs}
\usepackage{algorithm}
\usepackage{algorithmicx}
\usepackage{amsmath,amsfonts}
\usepackage{dsfont}
\usepackage{algpseudocode}
\usepackage{xcolor}
\usepackage{lipsum}
\usepackage{chngcntr}
\usepackage{tocloft}
\usepackage{pifont}
\usepackage{amssymb}
\usepackage[margin=2.90cm]{geometry}
\usepackage{rotating}
\usepackage{tikz}
\usepackage{adjustbox}
\usepackage{soul}

\makeatletter
\def\ps@pprintTitle{%
   \let\@oddhead\@empty
   \let\@evenhead\@empty
   \let\@oddfoot\@empty
   \let\@evenfoot\@oddfoot
}
\makeatother

\journal{}

\begin{document}

\begin{frontmatter}

\title{Mid-price Prediction Based on Machine Learning Methods with Technical and Quantitative Indicators}

\author[first]{Adamantios Ntakaris\corref{mycorrespondingauthor}}
\cortext[mycorrespondingauthor]{Corresponding author}
\ead{\string\href{mailto:adamantios.ntakaris@tut.fi}{adamantios.ntakaris@tuni.fi}}

\author[first]{Juho Kanniainen}

\author[first]{Moncef Gabbouj}

\author[second]{Alexandros Iosifidis}

\address[first]{Faculty of Information Technology and Communication Sciences, Tampere University, FI-33720, Tampere, Finland}
\address[second]{Department of Engineering, Electrical and Computer Engineering, Aarhus University, 8000, Aarhus, Denmark}

\begin{abstract}
Stock price prediction is a challenging task, but machine learning methods have recently been used successfully for this purpose. In this paper, we extract over 270 hand-crafted features (factors) inspired by technical and quantitative analysis and tested their validity on short-term mid-price movement prediction. We focus on a wrapper feature selection method using entropy, least-mean squares, and linear discriminant analysis. We also build a new quantitative feature based on adaptive logistic regression for online learning, which is constantly selected first among the majority of the proposed feature selection methods. This study examines the best combination of features using high frequency limit order book data from Nasdaq Nordic. Our results suggest that sorting methods and classifiers can be used in such a way that one can reach the best performance with a combination of only very few advanced hand-crafted features.
\end{abstract}

\begin{keyword}
high-frequency trading, mid-price, machine learning, technical analysis, quantitative analysis
\end{keyword}

\end{frontmatter}


\section{Introduction}

The problem under consideration in this paper is the prediction of a stock's mid-price movement during high-frequency financial trading. At a given time instance, the mid-price of a stock is defined as the average of the best ask and bid prices. We consider the mid-price as vital information for market makers who continuously balance inventories as well as for traders who need to be able to predict market movements in the correct direction to make money. Moreover, the mid-price facilitates the process of monitoring the markets' stability (i.e. spoofing identification).

\begin{figure*}[!t]
\centering
\includegraphics[scale=0.5]{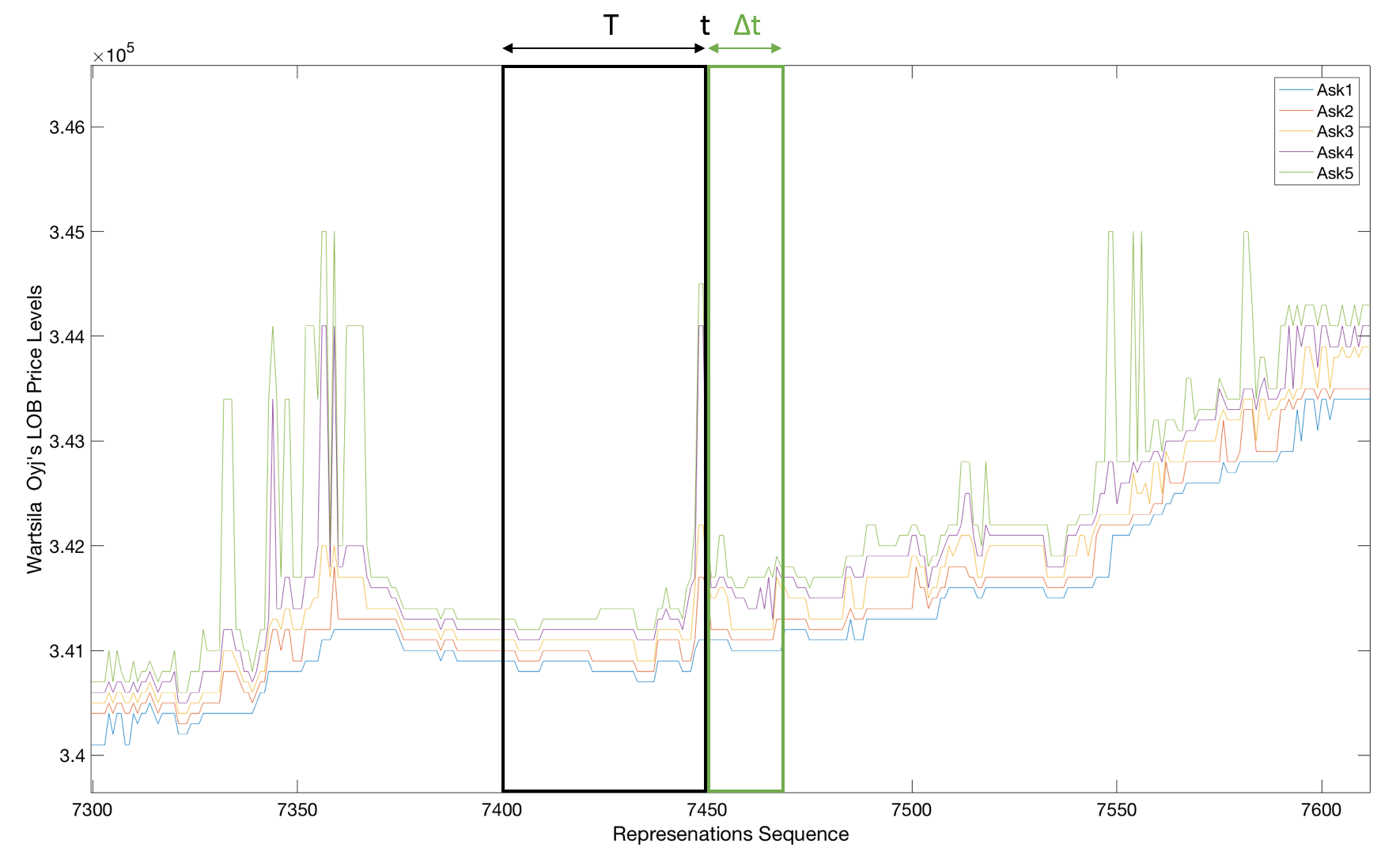}
\caption{The concept of mid-price prediction can be described as follows: at a given time instance $t$, the state of the stock is encoded in a vector-based representation calculated using a multi-dimensional time series information from a short-term time window of length $T$. Given this representation, the direction of the mid-price is predicted at a horizon of $\Delta t$.}
\label{fig:TimeSeriesPlot}
\end{figure*}

Over the past few years, several methods, such as those described in \cite{DASH201642}, \cite{gould2013limit},  \cite{passalis2017time}, \cite{sirignano2016deep}, \cite{thanh2017tensor}, \cite{tsantekidis2017forecasting}, and \cite{tsantekidis2017using}, have been proposed for analyzing stock market data. All these methods follow the standard classification pipeline formed by two processing steps. Given a time instance during the trading process, the state of the market is described based on a (usually short) time window preceding the current instance. A set of hand-crafted features is selected to describe the dynamics of the market, leading to a vector representation. Based on such a representation, a classifier is then employed to predict the state of the market at a time instance within a prediction horizon, as illustrated in \hyperref[fig:TimeSeriesPlot]{Fig.1}.

The majority of the studies as we see \hyperref[SS:MLHFT]{Section} \ref{SS:MLHFT} utilize a very limited amount of features without providing any motivation why they selected them, while here we cover the majority of: i) the technical indicators, ii) state-of-the-art limit order book (LOB) features, and iii) quantitative indicators. A work which also utilizes these sets of features can be found in \cite{ntakaris2019feature}.  Additionally, we propose a new advance quantitative feature that is selected first among several feature selection mechanisms for the task of mid-price movement prediction. The use of different hand-crafted features leads to encoding different properties of the financial time-series, and excluding some of these features can result in failing to exploit the relevant information. The definition of a good set of features is directly connected to the performance of the subsequent analysis, since any discarded information at this stage cannot be recovered later by the classifier.

One of the most widely followed approaches used to address this problem is using feature selection methods (e.g., \cite{Chandrashekar:2014:SFS:2577586.2577699}, \cite{MIAO2016919}) which can be performed in a wrapper fashion using various types of criteria for feature ranking. While the use of transformation-based dimensionality reduction techniques such as principal component analysis or linear discriminant analysis can lead to a similar processing pipeline, in this paper, we are interested in defining the set of features that convey most of the information in the data. The use of feature selection using unsupervised criteria, and in particular, the maximum entropy criterion, has been used in \cite{Battiti:1994:UMI:2325860.2328393} and \cite{LIU20091330}. The motivation behind this approach is the fact that as the entropy of a feature increases (when it is calculated in a set of data), the data variance and, thus, the information it encodes, also increases. However, the combination of many high-entropy features in a vector-based representation does not necessarily lead to good classification performance. This is because different dimensions of the adopted data representation need to encode different information.

In this paper, we provide an extensive analysis of various hand-crafted features (273 in total) for mid-price movement prediction. We base our analysis on a wrapped-based feature selection method (i.e., \cite{KOHAVI1997273}) that exploits unsupervised and supervised criteria for feature ranking. More specifically, we use maximum entropy (i.e., \cite{Battiti:1994:UMI:2325860.2328393}, \cite{LIU20091330}), maximum class discrimination based on linear discriminant analysis (LDA) \citep{Song:2010:FSB:1975506.1976555}, and regression-based classification \citep{Chen2011}. These different realizations of the feature selection method are applied to a wide pool of hand-crafted features. The list of hand-crafted features used in our study is selected to cover both basic and advanced features from two different trading approaches, which means those focusing on technical and quantitative analyses. Technical analysis is based on the fact that price prediction can be achieved by monitoring price and volume charts, while quantitative analysis focuses on statistical models and parameter estimation.

For the technical indicators, we calculate basic and advanced features accompanied by digital filters, while for the quantitative indicators, we primarily focus on time series analysis and online machine learning. We provide the full feature list and description and use it as input in twelve feature selection models (each corresponding to a different criterion and classifier combination) for our classification task. We not only present the best subset combination of these two types of features but also make a clear comparison of the two trading styles of feature tanks in terms of F1 performance (i.e. F1 score is a test to measure performance and is calculated as the harmonic mean of precision and recall). To the best of our knowledge this is the first study to define which types of information needs to be used for high-frequency time series description and classification.

The main contribution of our work lies on three pillars. The first pillar refers to the utilization of the majority of the technical indicator for the first time in the literature in the high-frequency trading universe. The second pillar refers to development of new quantitative feature, named adaptive logistic regression feature,  which selected first among several feature selection metrics. The third pillar, finally, refers to a fair and extent evaluation of these three feature sets (i.e., technical, quantitative, and LOB indicators) via the conversion of entropy, LDA, and LMS as feature selection criteria. This evaluation utilizes LMS, LDA, and radial basis function network (RBFN) as classifiers for the task of mid-price movememt prediction task. Our findings suggest that the best performance is reached by utilizing only very few, advanced, features derived from both quantitative and technical hand-crafted feature sets.

The remainder of the paper is organized as follows. We provide a comprehensive literature review in \hyperref[SS:MLHFT]{Section 2}. The problem statement and data description are provided in \hyperref[SS:Pr]{Section 3}. The list of hand-crafted features follows in \hyperref[SS:FT]{Section 4}. In \hyperref[SS:FS]{Section 5}, we describe the various realizations of the wrapper method adopted in our analysis, while \hyperref[SS:Re]{Section 6} provides empirical results, and \hyperref[SS:Con]{Section 7} concludes the paper. A detailed description of all features used in our experiments, as well as all ranking lists for each method, are provided in the Appendix section.

\section{Related Literature}\label{SS:MLHFT}
\noindent The rise of algorithmic trading, a type of trading requiring the use of computers under specific rules that can rapidly perform accurate calculations, suggests signal and statistical analyses. Several tools are based on these two types of analysis that a machine learning (ML) trader can utilize to select the best trade. However, which indicator or indicators (i.e. features) should be considered for a ML trader to secure a profitable move? Do historical and present prices contain all the relevant information? Finding answers to these questions is challenging due to the use of technical and quantitative analysis. The former category suggests that there is hidden information and patterns that can be extracted from historical data, whereas the latter suggests that statistical models and probabilities can provide relevant information to an ML trader.

Technical analysis (i.e., \cite{murphy1999technical}) has traditionally received less academic scrutiny than quantitative analysis. Nevertheless, several studies employ technical indicators as the main mechanism for signal analysis and price prediction. In the sphere of HFT, authors in \cite{pub:31778} utilize seven trading rule families as a measure of the impact of trading speed, while in \cite{kablan2010high} a fuzzy momentum analysis based on technical indicators for high speed trading is presented. Grammatical evolution is used in the E-mini S\&P 500 index futures market along with technical indicators for entry and exit trading exploration in \cite{6924111}. In \cite{lo2000foundations} authors provide an extensive investigation of charting analysis of nonparametric kernel regression for Nasdaq stocks via an automated strategy. A decision support system based on artificial neural networks (ANN) where six basic technical indicators are used as input features for signal generation is utilized in \cite{DASH201642}. An adaptive neuro fuzzy inference system (ANFIS) is used in \cite{kablan2009adaptive} for the FOREX market where technical indicators are utilized to benchmark ANFIS performance. Technical indicators are the basis for works in \cite{6557780}, \cite{4182391},  \cite{TEIXEIRA20106885}, and \cite{RODRIGUEZGONZALEZ201111489} for passive trading strategies (i.e. buy-and-hold) and stop-loss/stop-gain strategies. A list of ten technical indicators are utilized in \cite{PATEL2015259} as input features for several ML algorithms (i.e. ANN, SVM, random forest, and Naive Bayes) to predict stock trends. The interested reader can also find the implementation of ML methods with technical indicators in \cite{DEOLIVEIRA20137596}, \cite{935088}, \cite{DBLP:journals/corr/KhaidemSD16}, \cite{7850017}, \cite{WEN20101015}, and \cite{6643996}. Technical indicators are also used by \cite{baetje2016equity} for equity premium prediction in the US market, where they have been proved to be efficient in the out-of-sample period (1966 - 2014).
For the German bond market, authors in \cite{batchelor2007judgemental} extend the judgemental bootstrapping domain for the technical analysts' case. The authors prove that technical analysts can be as profitable as the statistical models of experts, by using only a subset of technical indicators. 

However, there is also quantitative analysis, which involves ML traders using complex mathematics and statistics as indicators when making trading decisions. Quantitative finance is a broad field that varies from topics like portfolio optimization (e.g., \cite{MAFI:MAFI179}, \cite{doi:10.1093/rfs/12.5.937}, \cite{INUIGUCHI200083}, \cite{10.2307/2975974}, \cite{markowitz1968portfolio}, \cite{doi:10.1287/mnsc.30.10.1143}) and asset pricing (e.g., \cite{fama1968risk}, \cite{french2003treynor}, \cite{jensen1972capital}, \cite{lintner1965valuation}, \cite{mossin1966equilibrium}, \cite{ross1977capital}, \cite{sharpe1964capital}) risk management (e.g., \cite{doi:10.1162/003465399558526}, \cite{JACF:JACF33}, \cite{hampton1982modern}, \cite{JACF:JACF27}), and time series analysis (e.g., \cite{10.2307/1925546}, \cite{box2015time}, \cite{iosifidis2012multidimensional}, \cite{taylor2008modelling}). In this work, we focus on time series analysis and use ideas from financial quantitative time series analysis that have been adjusted to ML. For example, authors in \cite{shen2012stock} use SVMs and decision trees via correlation analysis for stock market prediction. Another aspect of quantitative analysis is building trading strategies such as mean-reversion (i.e., \cite{poterba1988mean}). A simplistic example of this trading strategy is when a ML trader calculates Bollinger bands to spot trading signals and test a hypothesis. Furthermore, a financial time series is used for entry and exit signal exploration generated by Bollinger bands as described in \cite{LUBNAU2015312}. A time series analysis should also be tested for cointegration as suggested in \cite{engle1987co}. An additional aspect of quantitative analysis is the calculation of order book imbalance for order imbalance strategies. This idea is used as a feature in a deep neural network in \cite{sirignano2016deep}.

In the present work, we focus on extracted hand-crafted features based on technical and quantitative analysis. We show that a combination of features derived from these groups is able to improve forecasting ability. A combined method is employed by \cite{fang2003predictability} for asset returns predicatibility based on technical indicators and time series models. To the best of our knowledge this is the first attempt at a comparison between these trading schools using several feature selection methods in a wrapper fashion in the HFT literature.

\section{Problem Statement}\label{SS:Pr}
\noindent HF-type trading requires the constant analysis of market dynamics. One way to formulate these dynamics is constructing a limit order book (LOB), as illustrated in \hyperref[tab:LOB]{Table 2}. LOB is the cumulative order flow representing valid limit orders that are not executed nor cancelled, which are listed in the so-called message list, as illustrated in \hyperref[tab:Mes]{Table 1}. LOBs are multi-dimensional signals described by stochastic processes, and their dynamics are described as c\`adl\`ag functions(i.e., \cite{gould2013limit}).  Functions are formulated for a specific limit order (i.e. an order with specific characteristics in terms of price and volume at a specific time \textit{t}), as: $order$ = ($\textit{t}, Price_t, Volume_t $) that becomes active at time \textit{t} holds that: $order \in \mathcal{L}(\textit{t}), order\notin lim_{order' \uparrow order_x}\mathcal{L}(order')$.

\begin{table}[h!]\label{tab:Mes}
\centering
\scalebox{0.9}{
\begin{tabular}{cccccc}
\hline
Timestamp     & Id 	 & Price  & Quantity   & Event & Side \\
\hline
1275377039033 & 1372349 & 341100 & 300 & Submission & Bid \\
1275377039033 & 1372349 & 341100  &300  & Cancellation & Bid \\
1275377039037 & 1370659 & 343700 &100  & Submission & Ask \\
1275377039037 & 1370659 & 343700 &100  & Cancellation & Ask \\
1275377039037 & 1372352 & 341700 &150 & Submission & Bid \\
1275377039037 & 1372352 & 341700 & 150  & Cancellation & Bid \\
\hline
\end{tabular}}
\caption{Message list example. A sample from Wartsila Oyj on 01 June 2010.}
\label{tab:mb_example}
\end{table}

\begin{table}[h!]\label{tab:LOB}
\centering
\scalebox{0.8}{
\begin{tabular}{ccccccc}
& & \multicolumn{4}{c}{Level 1} & ...\\
 \cmidrule(l){2-7}
& & \multicolumn{2}{l}{Ask} & \multicolumn{2}{l}{Bid}   \\
\cmidrule(l){2-3} \cmidrule(l){4-5}
Timestamp      & Price & Quantity & Price & Quantity & \\
\cmidrule(l){1-3}  \cmidrule(l){4-5}
1127537703903 & 343600 & 100 & 342300 & 485 &  ... \\
1275377039033  & 343600 & 100 & 342300 & 485 & ... \\
1275377039037  & 343600 & 200 & 342300 & 485 &  ... \\
1275377039037  & 343600 & 200 & 342300 & 485 &  ... \\
1275377039037  & 343600 & 200 & 342300 & 485 &  ... \\
1275377039037  & 343600 & 200 & 342300 & 485 &  ... \\
1275377039037  & 343600 & 200 & 342300 & 485 &  ... \\
1275377039037  & 343600 & 200 & 342300 & 485 &  ... \\
\hline
\end{tabular}}
\caption{Order book example. A sample from Wartsila Oyj on 01 June 2010.}
\label{tab:ob_example}
\end{table}

Depending on how the LOB is constructed, we treat the new information according to event arrivals. The objective of our work is to predict the direction (i.e. up, down, and stationary condition) of the mid-price (i.e. $(p_a+p_b)/2$, where $p_a$ is the ask price and $p_b$ is the bid price at the first level of LOB). The goal is to utilize informative features based on the order flow (i.e. message list or message book [MB]) and LOB, which will help an ML trader improve the accuracy of mid-price movement prediction.

\section{Feature Pool}\label{SS:FT}
\noindent LOB and MB are the sources that we utilize for feature extraction. We provide the complete list of the features that have been explored in the literature for technical and quantitative trading in \hyperref[tab:flist]{Table 3}. The motivation for choosing the suggested list of features is based on an examination of all the basic and advanced features from technical analysis and comparisons with advanced statistical models, such as adaptive logistic regression for online learning. The present research has identified a gap in the existing literature concerning the performance of technical indicators and comparisons with quantitative models. The present work sets the groundwork for every future work in this direction since it provides insight into the features that are likely to achieve a high rank on the ordering list in terms of predictability power. To this end, we divide our feature set into three main groups. The first group of features is extracted according to \cite{doi:10.1080/14697688.2015.1032546} and \cite{DBLP:journals/corr/NtakarisMKGI17}. This group of features aims to capture the dynamics of LOB. This is possible if we consider the actual raw LOB data and relative intensities of different look-back periods of the trade types (i.e. order placement, execution, and cancellation). The second group of features is based on technical analysis. The suggested list describes most of the existing technical indicators (basic and advanced). Technical indicators might help traders spot hidden trends and patterns in their time series. The third group is based on quantitative analysis, which is mainly based on statistical models; it can provide statistics that are hidden in the data. This can be verified by the ranking process, where the advanced online feature (i.e. adaptive logistic regression) is placed first in most of the feature selection metrics (i.e. four out of five feature lists). The suggested features are fully described in \hyperref[SS:Feat]{Appendix A} apart from the description of the newly introduced adapative logistic regresion feature which follows.

\begin{table}
\centering
\scalebox{0.67}{
\begin{tabular}{rl}
\hline
\hline \\
\textbf{Feature Sets} & \textbf{Description} \\
\hline \\
First group:

& \\
\cline{1-1}

Basic 				&	n levels of LOB Data\\
\\
Time-Insensitive		&	Spread \& Mid-Price\\
					& 	Price Differences\\
					& 	Price \& Volume Means\\

					& 	Accumulated Differences\\
\\					
Time-Sensitive  		&	Price \& Volume Derivation\\
					&	Average Intensity per Type\\
					&	Relative Intensity Comparison\\
					&	Limit Activity Acceleration\\					
\hline \\

Second group: \\Technical Analysis 				& \\
\cline{1-1}
& Accumulation Distribution Line\\
& Awesome Oscillator\\
& Accelerator Oscillator\\
&Average Directional Index\\
&Average Directional Movement Index Rating\\
&Displaced Moving Average based on Williams Alligator Indicator\\
&Absolute Price Oscillator\\
&Aroon Indicator\\
&Aroon Oscillator\\
&Average True Range\\
&Bollinger Bands\\
&Ichimoku Clouds\\
&Chande Momentum Oscillator\\
&Chaikin Oscillator\\
&Chandelier Exit\\
&Center of Gravity Oscillator\\
&Donchian Channels\\
&Double Exponential Moving Average\\
&Detrended Price Oscillator\\
&Heikin-Ashi\\
&Highest High and Lowest Low\\
&Hull MA\\
&Internal Bar Strength\\
&Keltner Channels\\
&Moving Average Convergence/Divergence Oscillator\\
&Median Price\\
&Momentum\\
&Variable Moving Average\\
&Normalized Average True Range\\
&Percentage Price Oscillator\\
&Rate of Change\\
&Relative Strength Index\\
&Parabolic Stop and Reverse\\
&Standard Deviation\\
&Stochastic Relative Strength Index\\
&T3-Triple Exponential Moving Average\\
&Triple Exponential Moving Average\\
&Triangular Moving Average\\
&True Strength Index\\
&Ultimate Oscillator\\
&Weighted Close\\
&Williams \%R\\
&Zero-Lag Exponential Moving Average\\
&Fractals\\
&Linear Regression Line\\
&Digital Filtering: Rational Transfer Function\\
&Digital Filtering: Savitzky-Golay Filter\\
&Digital Filtering: Zero-Phase Filter\\
&Remove Offset and Detrend\\
&Beta-like Calculation\\
\hline \\
Third group: \\Quantitative Analysis 				& \\
\cline{1-1}
&Autocorrelation\\
&Partial Correlation\\
&Cointegration based on Engle-Granger test\\
&Order Book Imbalance\\
&\textbf{Adaptive Logistic Regression}\\
\hline
\hline
\end{tabular}}
\caption{Feature list for the three groups (description of the majority of the hand-crafted is in \hyperref[SS:Feat]{Appendix A} where the description of the newly introduced feature (in bold) can be found in \hyperref[sec:On]{Section} \ref{sec:On}).}
\label{tab:flist}
\end{table}

\subsection{Adaptive Logistic Regression}\label{sec:On}

\noindent We build a logistic regression model that we use as a feature in our experimental protocol. Motivation for this model is \cite{ng2000cs229} and \cite{sirignano2016deep} where the focal point is the local behavior of LOB levels. We extend this idea by doing online learning with an adaptive learning rate. More specifically, we use the Hessian matrix as our adaptive rate. We also report the ratio of the logistic coefficients based on the relationship of the LOB levels close to the best LOB level and the ones which are deeper in LOB.  Since $0\leqslant h_{\theta}(V) \leqslant 1$ and $V$ are the stock volumes for the first best six levels of the LOB, we formulate the model as follows:

\begin{equation}
h_{\theta}(V) = \dfrac{1}{1+e^{-\theta^{T}V}}
\end{equation}
be the logistic function (i.e. Hypothesis function) and $\theta^TV$ = $\theta_0 + \sum\limits_{j = 1}^{n}\theta_jV_j$. Parameter estimation is considered by calculating the parameter�s likelihood:
\begin{equation}
L(\theta) =\displaystyle \prod_{i =1}^{m}(h_{\theta}(V^{(i)}))^{y^{(i)}}(1 - h_{\theta}(V^{(i)}))^{1 - y^{(i)}}
\end{equation} for $m$ training samples and the cost function, based on this probabilistic approach, is as follows:
\begin{equation}
J(\theta) = \frac{1}{m}\sum\limits_{i =1}^{m}\big[-y^{(i)}log (h_{\theta}(V^{(i)})) - (1 - y^{(i)})log(1-h_{\theta}(V^{(i)}))\big].
\end{equation}
The next step is the process of choosing $\theta$s for optimizing (i.e. minimizing) J($\theta$). To do so, we will use Newton's update method:
\begin{equation}
\theta^{(s + 1)} = \theta^{(s)} - H^{-1}\nabla_{\theta}J,
\end{equation}
where the gradient is: $\nabla_{\theta}J$ = $\frac{1}{m}\sum\limits_{1}^{m}(h_{\theta}(V^{(i)}) - y^{(i)})V^{(i)}$
and the Hessian matrix is: $H$ = $\frac{1}{m}\sum\limits_{i = 1}^{m}\big[h_{\theta}(V^{(i)})\big(1 - h_{\theta}(V^{(i)})\big)V^{(i)}(V^{(i)})^T\big]$ with $V^{(i)}(V^{(i)})^T$ $\in \mathbb{R}^{(n+1) \times (n+1)}$ and $y^{(i)}$ are the labels which are calculated as the differences of the best level's ask (and bid) prices. The suggested labels describe a binary classification problem since we consider two states, one for change in the best ask price and another on for no change in the best ask price.

We perform the above calculation in an online manner. The online process considers the $9^{th}$ element of every 10 MB block multiplied by the $\theta$ coefficient first-order tensor to obtain the probabilistic behavior (we filter the obtained first-order tensor through the hypothesis function) of the $10^{th}$ event of the 10 MB block. The output is the feature representation expressed as scalar (i.e. probability) of the bid and ask price separately.

\section{Wrapper Method of Feature Selection}\label{SS:FS}
\noindent Feature selection is an area which focuses on applications with multidimensional datasets. An ML trader performs feature selection for three primary reasons: to reduce computational complexity, to improve performance, and to gain a better understanding of the underlying process. Feature selection, as a pre-processing method, can enhance classification power by adding features that contain information relevant to the task at hand. There are two metaheuristic feature selection methods: the wrapper method and the filter (i.e. transformation-based) method. We choose to perform classification based on the wrapper method since it considers the relationship among the features while the filter methods do not.

Our wrapper approach consists of five different feature subset selection criteria based on two linear and one non-linear methods for evaluation (see \hyperref[SS:Algo]{Algorithm 1} for a general description). More specifically, we convert entropy, least-mean-square (LMS), and  linear discriminant analysis (LDA) as feature selection criteria. For the last two cases (i.e., LMS and LDA) we provide two different selection criteria as follows: i) for LMS the first metric follows the $\mathcal{L}_2$-norm and the second metric is a statistical bias measure, and ii) for LDA the first metric is based on the ratio of the $within$-class scatter matrix while the second metric is derived according to the $between$-class scatter matrix. For classification evaluation we utilize LMS, LDA and a radial basis function network (i.e., \cite{broomhead1988radial}) as this utilized by \cite{iosifidis2015kernel} and \cite{DBLP:journals/corr/NtakarisMKGI17}). Our choice to apply these linear and non-linear classifiers is informed by the amount of data in our dataset (details will be provided in the following section). We measure classification performance according to accuracy, precision, recall, and the F1 scores for every possible combination of the hand-crafted features by utilizing LMS, LDA, and RBFN.  F1 score is defined as the harmonic average of the recall and precision, where recall is defined as $TP/TP + FN$ and precision is defined as $TP/(TP + FP)$ for TP, FN, and FP be the true positives, false negatives, and false positives, respectively.

\begin{algorithm}\label{SS:Algo}
\caption{Wrapper-Based Feature Selection}
\begin{algorithmic}[1]
\Procedure{$l_{opt}=Feature\_Select$}{$X, labels, criterion$}
\State $l = [1:D]$
\State $l_{opt} = [ \ \ ]$
\State $X_{opt} = [ \ \ ], \ [D, N] = size(X)$

\For{\texttt{$d = 1:D$}}
        \State $crit\_list = [\ \ ]$

		\For{\texttt{$i = 1:D-d+1$}}
			\State{$curr\_X = [X_{opt} ; X(i ; :)]$}
			\State{$crit\_list[i] = crit(curr\_X)$}
		\EndFor

\State{$[best\_d, best\_crit] = opt(crit\_list)$}
\State{$l_{opt}[d] = best\_d$}
\State{$l[best\_d] = [ \ \ ]$}
\State{$X_{opt} = [X_{opt} ; X(best\_d ; :)]$}
\State{$X(best\_d, :) = [ \ \ ]$}
\EndFor

\EndProcedure
\end{algorithmic}
\end{algorithm}

\bigbreak
\subsection{Feature Sorting}
\begin{figure*}[h!t]
\centering
\includegraphics[scale=0.43]{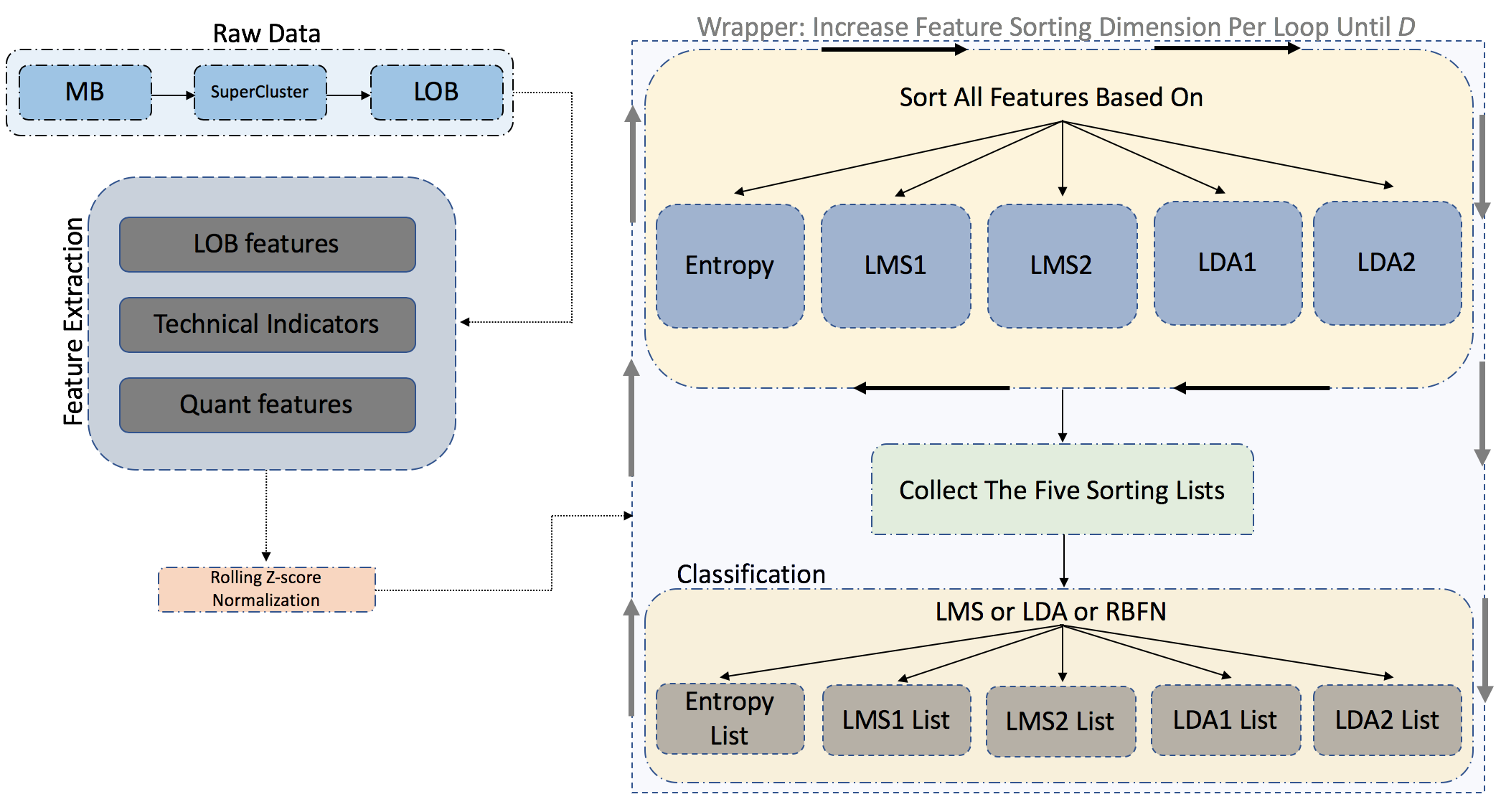}
\caption{The process of feature sorting and classification is based on the wrapper method. From left to right: 1) Raw data is converted to LOB data via supeclustering,  2) the feature extraction process follows (i.e., three feature sets are extracted), 3) then every feature set is normalized based on z-score in a rolling window basis (see \hyperref[SS:Re]{Section} \ref{SS:Re} for details), 4) Wrapper method: there are two main blocks in this process - the first block refers to the sorting (i.e., five different sorting criteria based on entropy, LMS1, LMS2, LDA1, and LDA2) of the feature sets independently (i.e., LOB features, technical indicators, and quantitative indicators are sorted separately) and all together (i.e., the three feature sets are merged and sorted all together) and the second block refers to the incremental classification (i.e., we increase the dimension of every sorting list during classification by one feature in every loop and average the final f1 scores per sorting list) based on three classifiers (i.e., LMS, LDA, and RBFN).}
\label{fig:Protocol}
\end{figure*}

\noindent We convert sample entropy, LMS, and LDA (for the latter two methods we use two different criteria for feature evaluation) into feature selection methods. A visual representation of the feature sorting method can be seen in \hyperref[fig:Protocol]{Fig.} \ref{fig:Protocol}.

\bigbreak
\subsubsection{Feature Sorting with Entropy}

\noindent We employ entropy \citep{richman2000physiological}, a measure of signal complexity where the signal is the time series of the multidimensional two-mode tensor with dimensions $\mathbb{R}^{p \times n}$, and where $p$ is the number of features and $n$ number of samples, as a measure of feature relevance. We calculate the bits of every feature in the feature set in an iterative manner and report the order. We measure the entropy as follows: $H(X) = -\sum\limits_{i = 1}^{p}p(x_i) \ log\ p(x_i)$, where $p(x_i)$ is the probability of the frequency per feature for the given data samples.

\subsubsection{Feature Sorting with Least-Mean-Square}\label{SS:LMS}
\noindent We perform feature selection based on the least-mean square classification rate (LMS1) and $\mathcal{L}_2$-norm (LMS2). LMS is a fitting method which aims to produce an approximation that minimizes the sum of squared differences between given data and predicted values. We use this approach to evaluate the relevance of our hand-crafted features. A hand-crafted feature evaluation is performed sequentially via LMS. More specifically, each of the features is evaluated based on the classification rate, the $\mathcal{L}_2$-norm of the predicted labels, and the ground truth. The evaluation process is performed as follows: $\textbf{H}\textbf{W} = \textbf{T}$, where $\textbf{H} \in \mathbb{R}^{p_i \times n}$ is the input data with feature dimension $p_i$ where it is calculated incrementally for the number of training samples $n$, $\textbf{W} \in \mathbb{R}^{p_i \times \#cl}$ are the weighted coefficients for the number of features $p_i$ of the number(\#) of classes (i.e. up, down, and stationary labelling), and $\textbf{T} \in \mathbb{R}^{\#cl \times n}$ represents the target labels of the training set. The weight coefficient matrix $\textbf{W}$ is estimated via the following formula: $\textbf{W}$ = $\textbf{H}^{\dagger}\textbf{T}$, where $\textbf{H}^{\dagger}$ is the Moore-Penrose pseudoinverse matrix.

\subsubsection{Feature Sorting with Linear Discriminant Analysis}\label{SS:LDA}
\noindent Linear discriminant analysis (LDA) can be used for classification and dimensionality reduction. However, instead of performing these two tasks, we convert LDA into a feature selection algorithm. We measure feature selection performance based on two metrics. One is the classification rate (LDA1) and the other is based on the error term (LDA2), which we define as the ratio of the \textit{within}-class scatter matrix and the \textit{between}-class scatter matrix. The main objective of LDA is finding the projection matrix $\textbf{W} \in \mathbb{R}^{m \times \#cl-1}$, where $m$ is the sample dimension, and $\#cl$ is the number of classes, such that $Y = W^TX$ maximizes the class separation. For the given sample set $X = X_1 \cup X_2 \cup ... \cup X_{\#cl}$, where $X_k = \{ x_1^k, ... , x_{\ell_k}^k \}_{k = 1, ..., \#cl}$ is the class-specific data subsample, we try to find $\textbf{W}$ that maximizes Fisher's ratio $J(\textbf{W})$ = $\dfrac{trace\left(\textbf{W}^T S_B \textbf{W}\right)}{trace\left(\textbf{W}^T S_W \textbf{W}\right)}$, where $S_B = \sum\limits_{i=1}^{\#cl}N_i(\mu_i - \mu)(\mu_i - \mu)^T$ and $S_W = \sum\limits_{i=1}^{C}\sum\limits_{x \in X_i}^{d}(x-\mu_i)(x-\mu_i)^T$ are the $between$-class and $within$-class scatter matrices, respectively, with $\mu_i = \frac{1}{\ell_k}\sum\limits_{k \in \#cl}^{}X_k$ and $\mu = \frac{1}{m} \sum\limits_{k \in \#cl}^{}\ell_kX_k$. In a similar fashion, we perform calculations for the projected samples $\textbf{y}$ with $\widetilde{\mu_i} = \frac{1}{\ell_k}\sum\limits_{k \in \#cl}^{}Y_k$ and $\widetilde{\mu} = \frac{1}{m}\sum\limits_{k \in \#cl}^{}\ell_k Y_k$, while the scatter matrices (i.e. $within$ and $between$ scatter matrices, respectively) are $\widetilde{S_W} = \sum\limits_{i=1}^{\#cl}\sum\limits_{k \in \#cl }{}(y - \tilde{\mu_i})(y - \tilde{\mu_i})^T$ and $\widetilde{S_B} = \sum\limits_{i = 1}^{\#cl}\ell_k(\widetilde{\mu_i} - \widetilde{\mu})(\widetilde{\mu_i} - \widetilde{\mu})^T$. The above calculations constitute the basis for the two metrics that we use to evaluate the hand-crafted features incrementally. The two evaluation metrics are based on the classification rate and the ratio of the $within$-class and $between$-class scatter matrices of the projected space $Y$.

\subsection{Classification for Feature Selection}
\noindent We perfrom classification evaluation based on three classifiers: LMS, LDA, and RBFN. Theory for the first two discussed in Sections \hyperref[SS:LMS]{} \ref{SS:LMS} and \hyperref[SS:LDA]{} \ref{SS:LDA} while RBFN classifiers is descibed in the following section.

\subsubsection{RBFN Classifier}\label{SS:SLFN}
\begin{figure}[h!]
\centering
\includegraphics[scale=0.30]{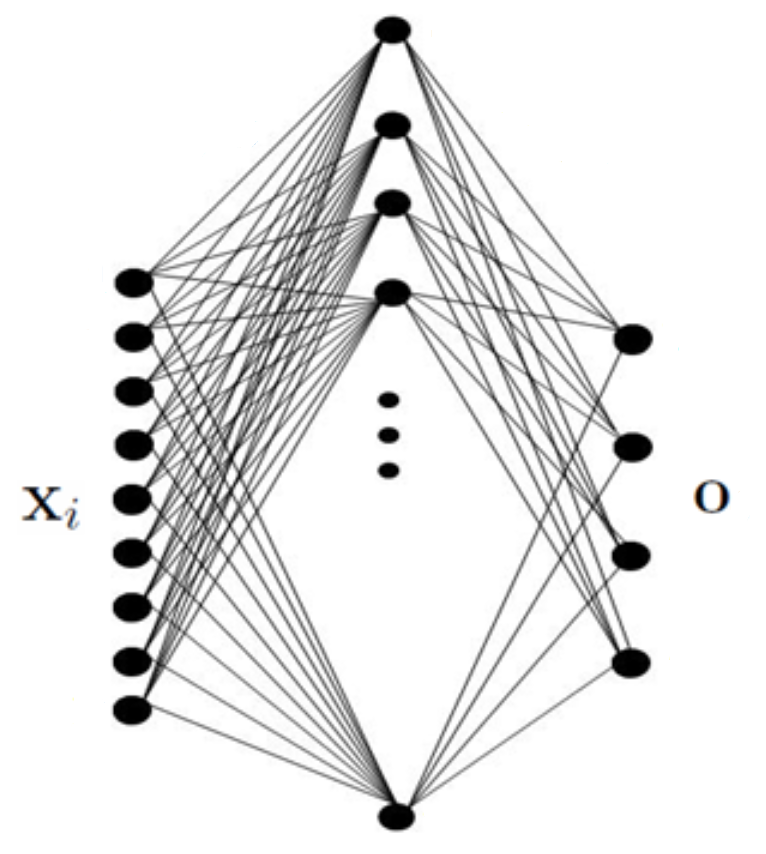}
\caption{SLFN}
\label{fig:SLFN}
\end{figure}

\noindent We utilize a SLFN (\hyperref[fig:SLFN]{Fig.} \ref{fig:SLFN}) as this suggested by 
\cite{huang2006extreme}. The description and the implememntation can be also found in \cite{DBLP:journals/corr/NtakarisMKGI17}. This type of model is formed as in \hyperref[fig:SLFN]{Fig.} \ref{fig:SLFN}.

In order to train fast this model we follow \cite{zhang2009prototype}, and \cite{iosifidis2017approximate}. We utilize $K$-means clustering for $K$ prototype vectors identification, which then are used as the network's hidden layer weights. Having identified the network's hidden layer weights $\mathbf{V} \in \mathbb{R}^{D \times K}$, the input data $\mathbf{x}_i, \:i=1,\dots,N$ are mapped to vectors $\mathbf{h}_i \in \mathbb{R}^K$ in a non-linear way, expressing the data representations in the feature space determined by the network's hidden layer outputs $\mathbb{R}^K$. Then radial basis function is utilized, i.e. $\mathbf{h}_i = \phi_{RBF}(\mathbf{x}_i)$, calculated in an element-wise manner, as follows:
\begin{equation}
h_{ik} = \exp{\left( \frac{ \|\mathbf{x}_i-\mathbf{v}_k\|^2_2 }{2 \sigma^2}  \right)}, \:\:k=1,\dots,K,
\end{equation}
where $\sigma$ is a hyper-parameter denoting the spread of the RBF neuron and $\mathbf{v}_k$ corresponds to the $k$-th column of $\mathbf{V}$. 

The network's output weights $\mathbf{W} \in \mathbb{R}^{K \times C}$ are determined by solving the following equation:
\begin{equation}\label{Eq:PVMcriterion}
\mathbf{W}^* = \underset{\mathbf{W}}{\arg min} \:\: \| \mathbf{W}^T \mathbf{H} - \mathbf{T} \|^2_F + \lambda \|\mathbf{W}\|^2_F,
\end{equation}
where $\mathbf{H} = [\mathbf{h}_1,\dots,\mathbf{h}_N]$ is a matrix formed by the network's hidden layer outputs for the training data and $\mathbf{T}$ is a matrix formed by the network's target vectors $\mathbf{t}_i, \:i=1,\dots,N$ as defined in \hyperref[SS:RR]{Section} \ref{SS:RR}. The network's output weights are given by:
\begin{equation}\label{Eq:PVMsolution2}
\mathbf{W} = \left(\mathbf{H} \mathbf{H}^T + \lambda \mathbf{I} \right)^{-1} \mathbf{H} \mathbf{T}^T.
\end{equation}
\noindent Then a new (test) sample $\mathbf{x} \in \mathbb{R}^D$ is mapped on its corresponding representations in spaces $\mathbb{R}^K$ and $\mathbb{R}^C$, i.e. $\mathbf{h} = \phi_{RBF}(\mathbf{x})$ and $\mathbf{o} = \mathbf{W}^T \mathbf{h}$, respectively. fFinally, the classification task is based on the maximal network output, i.e.:
\begin{equation}
l_{\mathbf{x}} = \underset{k}{\arg max} \:\:o_k.
\end{equation}
 
\section{Results}\label{SS:Re}
\noindent In this section, we provide details regarding the conducted experiments. The experiments are based on the idea of a mid-price prediction state (i.e. up, down, and stationary) for ITCH feed data in millisecond resolution. For the experimental protocol, we followed the setup in \cite{DBLP:journals/corr/NtakarisMKGI17}, which is based on the  anchored cross-validation format. According to this format, we use the first day as training and second day as testing for the first fold, whereas the second fold consists of the previous training and testing periods as a training set, and the next day is always used as a test set. Each of the training and testing sets contains the hand-crafted feature representations for all the five stocks from the FI-2010 dataset. Hence, we obtain a mode-three tensor of dimensions $273 \times 458,125$. The first dimension is the number of features, whereas the second one is the number of sample events. At this point, we must specify that the process of hand-crafted feature extraction in \hyperref[SS:FT]{Section 5} is conducted in the full length of the given information based on MB with 4,581,250 events. The motivation for taking separate blocks of messages of ten events is the time-invariant nature of the data. To keep the ML trader fully informed regarding MB blocks, we use features that convey this information by calculating, among others, averages, regression, risk, and online learning feedback.

The results we present here are the mid-price predictions for the next $10^{th}$, $20^{th}$, and $30^{th}$ events (i.e. translated into MB events) or else one, two, and three next events after the current state translated into a feature representations setup. The prediction performance of these events is measured by the accuracy, precision, recall and F1 score, whereas we emphasize the F1 score. We focus on the F1 score because it can only be affected in one direction by skewed distributions for unbalanced classes, as observed in our data. Performance metrics are calculated against the mid-price labelling calculation of ground truth extraction. More specifically, we extract labels based on the percentage change of the smoothed mid-price with a span window of 9, for our supervised learning methods, by calculating it as follows: $L_{1} = \dfrac{MP_{next} - MP_{curr}}{MP_{curr}}$, where $MP_{curr}$ is the current mid-price, and $MP_{next}$ is the next mid-price. We threshold the percentage change identification by a fixed number $\gamma = 0.002$ and perform a rolling z-score normalization on our dataset in order to avoid look-ahead bias\footnote{Look ahead bias refers to the process that future information is injected to the training set.}. The rolling window z-score normalization is based on the anchored cross validation setup which means that the normalization of the training set is totally unaffected from any fututre information.

\begin{figure}[H]
\centering
\includegraphics[scale=0.250]{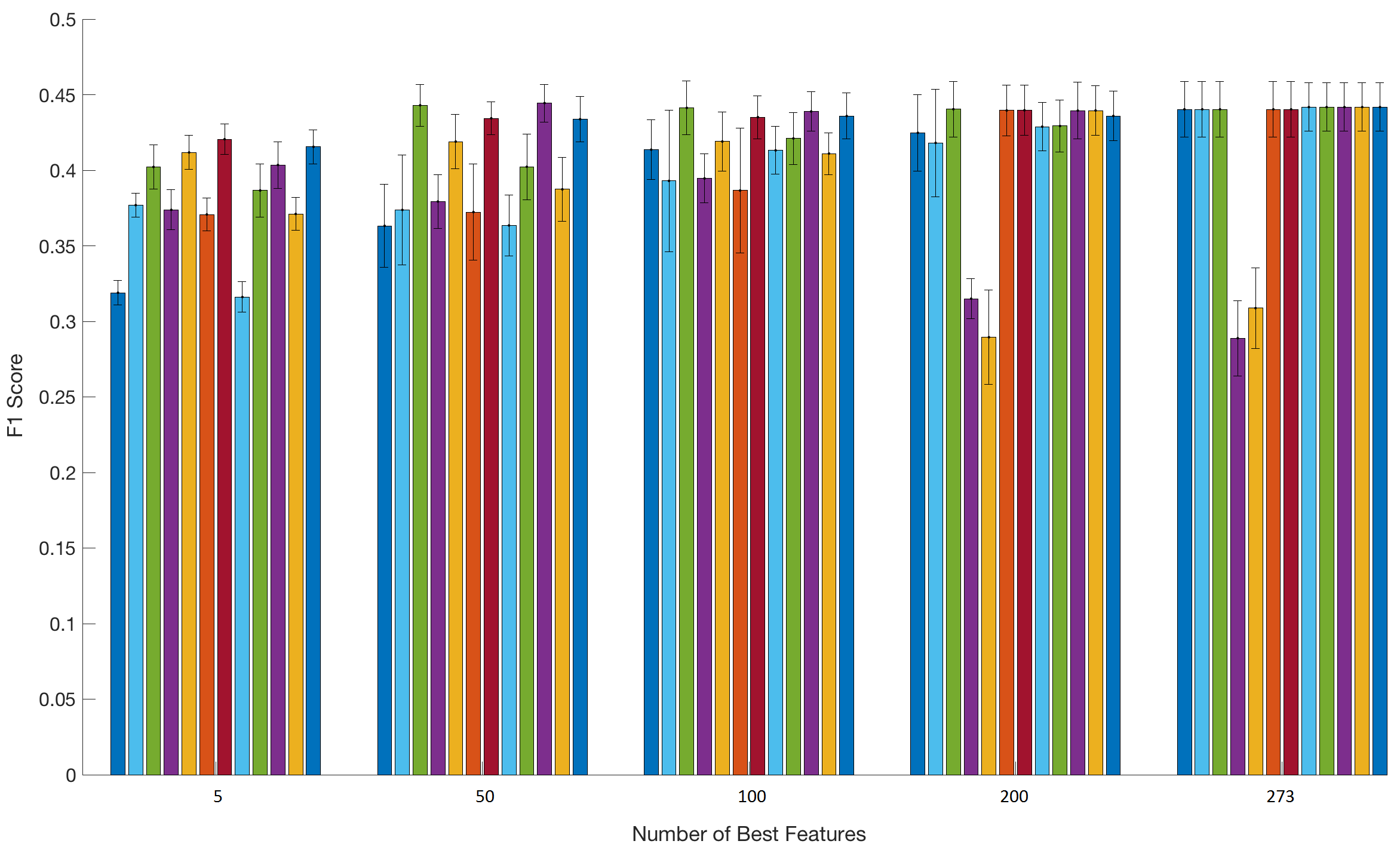}
\caption{Bar plots with variance present the average (i.e. average F1 performance for the 9-fold protocol for all the features) F1 score  of the 12 different models for the cases of 5, 50, 100, 200, and 273 number of best features. The order of the models from the left to the right column is (1) feature list sorted based on entropy and classified based on LMS, (2) feature list sorted based on LMS1 and classified based on LMS, (3) feature list sorted based on LMS2 and classified based on LMS, (4) feature list sorted based on LDA1 and classified based on LDA, (5) feature list sorted based on LDA2 and classified based on LDA, (6) feature list sorted based on LDA1 and classified based on LMS, (7) feature list sorted based on LDA2 and classified based on LMS, (8) feature list sorted based on LDA2 and classified based on LMS, (9) feature list sorted based on entropy and classified based on RBFN, (10) feature list sorted based on LMS2 and classified based on RBFN, (11) feature list sorted based on LDA1 and classified based on RBFN, and (12) feature list sorted based on LDA2 and classified based on RBFN.}
\label{fig:BarPlot}
\end{figure}

\begin{figure}[H]
\begin{center}
\hspace*{-0.6cm}
\centering
\begin{adjustbox}{addcode={\begin{minipage}{\width}}{\caption{%
      F1 performance per number of best features sequence for 10 events as the projected horizon, where lines 	represent the 5 different sorting methods as classified based on LMS, LDA, and RBFN.
      }\end{minipage}},rotate=90,center}
      \includegraphics[origin=c, scale=0.340]{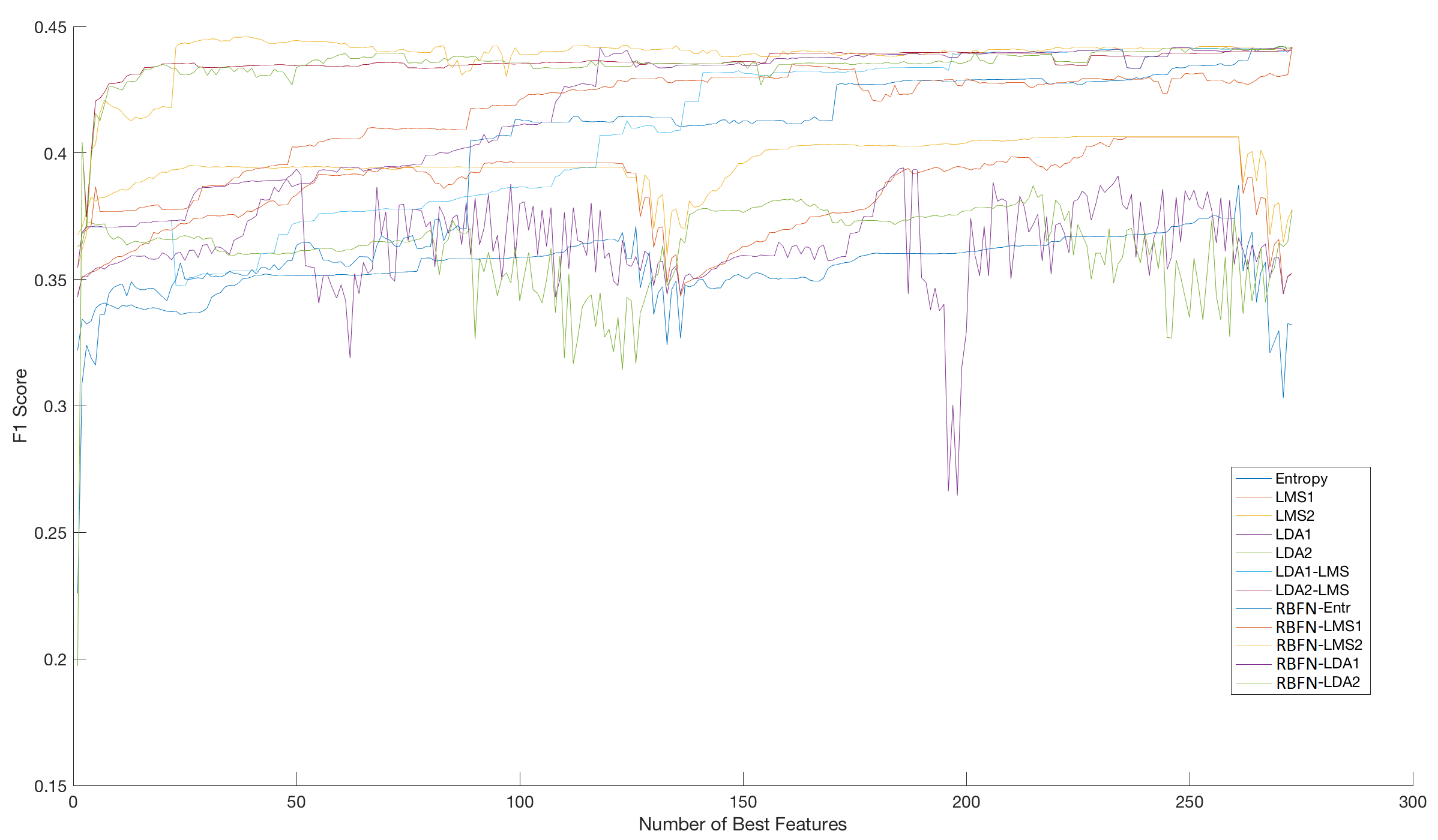}
  \end{adjustbox}

\label{fig:FeatureNumber}
\end{center}
\end{figure}

We report our results in \hyperref[tab:Table4]{Table 4} - \hyperref[tab:Table8]{Table 8} of the best feature selection list for the five sorting lists (i.e. based on entropy, LMS1, LMS2, LDA1, and LDA2) according to the supervised linear and non-linear classifiers of LMS, LDA, and RBFN. For the last classifier (i.e. RBFN), we use a multilayer perceptron based on the extreme learning machine model with a twist in the initialization process of weights calculation based on the k-means algorithm. The full description of this method can be found in \cite{DBLP:journals/corr/NtakarisMKGI17}. We provide results based on the whole feature pool (see \hyperref[tab:Table4]{Table 4}), the first feature pool according to \cite{doi:10.1080/14697688.2015.1032546} and \cite{DBLP:journals/corr/NtakarisMKGI17} (see \hyperref[tab:Table5]{Table 5}), based only on technical indicators (see \hyperref[tab:Table6]{Table 6}) and quantitative indicators (see \hyperref[tab:Table7]{Table 7}). More specifically, for the first feature pool, we have 135 features, while for the second pool we have 83 features, and for the last pool we have 55 features; in total, we have 273 features. The number of best features that used in the above methods is different in every case and can be monitored in \hyperref[fig:BarPlot]{Fig.2} and \hyperref[fig:FeatureNumber]{Fig.3}. We should point out that we tested all the possible combinations for the five sorting methods and the three classifiers (i.e., 15 different cases) but we report results that had some variations. For instance, in \hyperref[tab:Table4]{Table} \ref{tab:Table4} we report results for entropy as sorting method and LMS together with RBFN as classifiers but not with LDA (as classifier) since the last method reports similar results. 

\begin{table}[H]\label{tab:Table4}
\centering
\scalebox{0.797}{
\begin{tabular}{cccccccc}
\hline
Sorting & Classifier & T &$Accuracy$ & $Precision$ & $Recall$ & $F1$  \\
\hline
Entropy	& LMS	& 10	& 0.529 $\pm$ 0.059	& 	0.447 $\pm$ 0.007	&	0.477 $\pm$ 0.013	&	0.440 $\pm$ 0.018	\\

LMS1 	& LMS	& 10	& 0.540 $\pm$ 0.059	&	0.437 $\pm$ 0.007	&	0.456 $\pm$ 0.013	&	0.430 $\pm$ 0.018	\\

LMS2	& LMS 	& 10	& 0.538 $\pm$ 0.052	&	0.447 $\pm$ 0.005	&	0.478 $\pm$ 0.013	&	0.444 $\pm$ 0.011	\\

LDA1		& LDA	& 10	& 0.616 $\pm$ 0.048	&	0.408 $\pm$ 0.019	&	0.398 $\pm$ 0.011	&	0.397 $\pm$ 0.015	\\

LDA2	& LDA	& 10	& 0.543 $\pm$ 0.057	&	0.430 $\pm$ 0.010	&	0.455 $\pm$ 0.017	&	0.429 $\pm$ 0.018	\\

LDA1		& LMS	& 10	& 0.604 $\pm$ 0.068	&	0.468 $\pm$ 0.035	&	0.431 $\pm$ 0.042	&	0.408 $\pm$ 0.035	\\

LDA2	& LMS	& 10	& 0.522 $\pm$ 0.026	&	0.441 $\pm$ 0.020	&	0.473 $\pm$ 0.007	&	0.435 $\pm$ 0.007	\\

Entropy	& RBFN	& 10	& 0.474 $\pm$ 0.046	&	0.420 $\pm$ 0.031	&	0.445 $\pm$ 0.039	&	0.400 $\pm$ 0.039	\\

LMS1	& RBFN	& 10	& 0.600 $\pm$ 0.045	&	0.436 $\pm$ 0.019	&	0.425 $\pm$ 0.021	&	0.417 $\pm$ 0.019	\\

LMS2	& RBFN	& 10	& 0.537 $\pm$ 0.016	&	0.442 $\pm$ 0.011	&	0.470 $\pm$ 0.016	&	0.439 $\pm$ 0.012 \\

LDA1		& RBFN	& 10	& 0.585 $\pm$ 0.061	&	0.443 $\pm$ 0.018	&	0.438 $\pm$ 0.037	&	0.419 $\pm$ 0.026	\\

LDA2	& RBFN	& 10	& 0.528 $\pm$ 0.029	&	0.438 $\pm$ 0.020	&	0.467 $\pm$ 0.010	&	0.434 $\pm$ 0.017	\\
\hline
Entropy	& LMS	& 20	& 0.503 $\pm$ 0.049	& 	0.469 $\pm$ 0.008	&	0.482 $\pm$ 0.014	&	0.462 $\pm$ 0.017	\\

LMS1	& LMS	& 20	& 0.503 $\pm$ 0.049	&	0.470 $\pm$ 0.008	&	0.482 $\pm$ 0.014	&	0.462 $\pm$ 0.017	\\

LMS2	& LMS	& 20	& 0.503 $\pm$ 0.049	&	0.469 $\pm$ 0.008	&	0.481 $\pm$ 0.014	&	0.462 $\pm$ 0.018	\\

LDA1		& LDA	& 20	& 0.478 $\pm$ 0.060	&	0.400 $\pm$ 0.038	&	0.404 $\pm$ 0.041	&	0.393 $\pm$ 0.018	\\

LDA2	& LDA	& 20	& 0.505 $\pm$ 0.046	&	0.452 $\pm$ 0.009	&	0.461 $\pm$ 0.012	&	0.450 $\pm$ 0.012	\\

LDA1		& LMS	& 20	& 0.530 $\pm$ 0.032	&	0.457 $\pm$ 0.024	&	0.426 $\pm$ 0.048	&	0.401 $\pm$ 0.048	\\

LDA2	& LMS	& 20	& 0.499 $\pm$ 0.019	&	0.462 $\pm$ 0.019	&	0.476 $\pm$ 0.007	&	0.457 $\pm$ 0.015	\\

Entropy	& RBFN	& 20	& 0.464 $\pm$ 0.038	&	0.436 $\pm$ 0.033	&	0.448 $\pm$ 0.035	&	0.425 $\pm$ 0.036	\\

LMS1	& RBFN	& 20	& 0.519 $\pm$ 0.023	&	0.430 $\pm$ 0.016	&	0.417 $\pm$ 0.022	&	0.412 $\pm$ 0.027	\\

LMS2	& RBFN	& 20	& 0.508 $\pm$ 0.010	&	0.456 $\pm$ 0.015	&	0.466 $\pm$ 0.018	&	0.454 $\pm$ 0.017	\\

LDA1		& RBFN	& 20	& 0.523 $\pm$ 0.025	&	0.441 $\pm$ 0.024	&	0.429 $\pm$ 0.041	&	0.416 $\pm$ 0.046	\\

LDA2	& RBFN	& 20	& 0.502 $\pm$ 0.018	&	0.454 $\pm$ 0.019	&	0.465 $\pm$ 0.009	&	0.452 $\pm$ 0.015	\\
\hline
Entropy	& LMS	& 30	& 0.503 $\pm$ 0.042	& 	0.475 $\pm$ 0.013	&	0.484 $\pm$ 0.014	&	0.470 $\pm$ 0.019	\\

LMS1	& LMS	& 30	& 0.503 $\pm$ 0.042	&	0.475 $\pm$ 0.013	&	0.484 $\pm$ 0.014	&	0.470 $\pm$ 0.019	\\

LMS2	& LMS	& 30	& 0.503 $\pm$ 0.043	&	0.474 $\pm$ 0.012	&	0.482 $\pm$ 0.014	&	0.461 $\pm$ 0.019	\\

LDA1		& LDA	& 30	& 0.464 $\pm$ 0.048	&	0.414 $\pm$ 0.025	&	0.420 $\pm$ 0.027	&	0.403 $\pm$ 0.018	\\

LDA2	& LDA	& 30	& 0.500 $\pm$ 0.043	&	0.457 $\pm$ 0.012	&	0.464 $\pm$ 0.013	&	0.455 $\pm$ 0.014	\\

LDA1		& LMS	& 30	& 0.489 $\pm$ 0.018	&	0.451 $\pm$ 0.030	&	0.429 $\pm$ 0.051	&	0.405 $\pm$ 0.072	\\

LDA2	& LMS	& 30	& 0.496 $\pm$ 0.016	&	0.476 $\pm$ 0.018	&	0.479 $\pm$ 0.009	&	0.472 $\pm$ 0.015	\\

Entropy	& RBFN	& 30	& 0.464 $\pm$ 0.035	&	0.446 $\pm$ 0.035	&	0.449 $\pm$ 0.034	&	0.440 $\pm$ 0.037	\\

LMS1	& RBFN	& 30	& 0.471 $\pm$ 0.018	&	0.425 $\pm$ 0.018	&	0.414 $\pm$ 0.020	&	0.409 $\pm$ 0.026	\\

LMS2	& RBFN	& 30	& 0.494 $\pm$ 0.014	&	0.464 $\pm$ 0.021	&	0.466 $\pm$ 0.021	&	0.461 $\pm$ 0.024	\\

LDA1		& RBFN	& 30	& 0.481 $\pm$ 0.022	&	0.438 $\pm$ 0.034	&	0.428 $\pm$ 0.045	&	0.415 $\pm$ 0.057	\\

LDA2	& RBFN	& 30	& 0.493 $\pm$ 0.017	&	0.465 $\pm$ 0.018	&	0.467 $\pm$ 0.010	&	0.463 $\pm$ 0.016	\\
\hline
\end{tabular}}
\label{tab:Table1}
\caption{F1-macro (i.e. F1-macro = $\frac{1}{C}\sum_{k\in C}F1_k$, with $C$ as the number of classes for the 9-fold experimental protocol) results, based on the total feature pool, for the five sorting lists classified per LMS, LDA, and RBFN for the next $T=$$10^{th}$, $20^{th}$, and $30^{th}$ events, respectively, as the predicted horizon. The number of best features used in the above methods is different in every case (as seen in \hyperref[fig:FeatureNumber]{Fig. 3}). }
\end{table}

\begin{table}[H]\label{tab:Table5}
\centering
\scalebox{0.75}{
\begin{tabular}{cccccccc}
\hline
Sorting  & Classifier&T &$Accuracy$ & $Precision$ & $Recall$ & $F1$  \\
\hline
Entropy	&	LMS & 10	& 0.420 $\pm$ 0.025	& 	0.379 $\pm$ 0.011 &		0.397 $\pm$ 0.011	&	0.355 $\pm$ 0.013	\\

LMS1 	&	LMS & 10	& 0.574 $\pm$ 0.055	&	0.402 $\pm$ 0.013 &	0.396 $\pm$ 0.018	&	0.384 $\pm$ 0.020	\\

LMS2	&	LMS & 10	& 0.519 $\pm$ 0.015	&	0.400 $\pm$ 0.009 &	0.413 $\pm$ 0.009	&	0.396 $\pm$ 0.010	\\

LDA1		& 	LDA & 10	& 0.561 $\pm$ 0.090	&	0.389 $\pm$ 0.018 &	0.382 $\pm$ 0.016	&	0.363 $\pm$ 0.030	\\

LDA2	&	LDA & 10	& 0.507 $\pm$ 0.041	&	0.373 $\pm$ 0.014 &	0.384 $\pm$ 0.017	&	0.362 $\pm$ 0.019	\\
\hline
Entropy	&	LMS & 20	& 0.386 $\pm$ 0.018	& 	0.386 $\pm$ 0.015 &	0.397 $\pm$ 0.015	&	0.363 $\pm$ 0.016	\\

LMS1	&	LMS & 20	& 0.527 $\pm$ 0.027	&	0.411 $\pm$ 0.013 &		0.389 $\pm$ 0.015	&	0.375 $\pm$ 0.029	\\

LMS2	&	LMS & 20	& 0.462 $\pm$ 0.013	&	0.405 $\pm$ 0.013 &	0.410 $\pm$ 0.009	&	0.400 $\pm$ 0.012	\\

LDA1		&	LDA & 20	& 0.529 $\pm$ 0.031	&	0.406 $\pm$ 0.017	&	0.381 $\pm$ 0.011	&	0.360 $\pm$ 0.024	\\

LDA2	&	LDA & 20	&  0.461 $\pm$ 0.036	&	0.378 $\pm$ 0.016	&	0.380 $\pm$ 0.016	&	0.368 $\pm$ 0.021	\\
\hline
Entropy	&	LMS & 30	& 0.391 $\pm$ 0.016	& 	0.395 $\pm$ 0.018	&	0.401 $\pm$ 0.015	&	0.380 $\pm$ 0.017	\\

LMS1	&	LMS & 30	& 0.459 $\pm$ 0.025	&	0.405 $\pm$ 0.017	&	0.388 $\pm$ 0.020	&	0.366 $\pm$ 0.040	\\

LMS2	&	LMS & 30	& 0.432 $\pm$ 0.009	&	0.407 $\pm$ 0.015	&	0.409 $\pm$ 0.013	&	0.401 $\pm$ 0.016	\\

LDA1		&	LDA & 30	& 0.447 $\pm$ 0.041	&	0.391 $\pm$ 0.018	&	0.377 $\pm$ 0.017	&	0.352 $\pm$ 0.037	\\

LDA2	&	LDA & 30	& 0.418 $\pm$ 0.028	&	0.373 $\pm$ 0.016	&	0.375 $\pm$ 0.015	&	0.361 $\pm$ 0.018	\\
\hline
\end{tabular}}
\caption{F1-macro (i.e. F1-macro = $\frac{1}{C}\sum_{k\in C}F1_k$, with $C$ as the number of classes for the 9-fold experimental protocol) results, based only on the hand-crafted features from \cite{DBLP:journals/corr/NtakarisMKGI17}, for the five sorting lists classified based on LMS, LDA, and RBFN for the next $T=$$10^{th}$, $20^{th}$, and $30^{th}$ events, respectively, as the predicted horizon. The number of best features used in the above methods is different in every case (as seen in \hyperref[fig:FeatureNumber]{Fig. 3}).}
\end{table}

\begin{table}[H]\label{tab:Table6}
\centering
\scalebox{0.75}{
\begin{tabular}{cccccccc}
\hline
Sorting  & Classifier & T & $Accuracy$ & $Precision$ & $Recall$ & $F1$  \\
\hline
Entropy	&	LMS & 10	& 0.456 $\pm$ 0.038	& 	0.372 $\pm$ 0.021	&	0.380 $\pm$ 0.014	&	0.353 $\pm$ 0.020	\\

LMS1 	&	LMS & 10	& 0.497 $\pm$ 0.066	&	0.371 $\pm$ 0.017	&	0.377 $\pm$ 0.021 &	0.354 $\pm$ 0.024	\\

LMS2	&	LMS & 10	& 0.460 $\pm$ 0.016	&	0.383 $\pm$ 0.012	&	0.394 $\pm$ 0.009 &	0.365 $\pm$ 0.010	\\

LDA1		& 	LDA & 10	& 0.517 $\pm$ 0.064	&	0.367 $\pm$ 0.015	&	0.371 $\pm$ 0.020 &	0.344 $\pm$ 0.026	\\

LDA2	&	LDA & 10	& 0.475 $\pm$ 0.023	&	0.371 $\pm$ 0.010	&	0.382 $\pm$ 0.009	&	0.351 $\pm$ 0.015	\\
\hline
Entropy	&	LMS & 20	& 0.430 $\pm$ 0.029	& 	0.384 $\pm$ 0.025 &	0.387 $\pm$ 0.017	&	0.371 $\pm$ 0.023 \\

LMS1	&	LMS & 20	& 0.480 $\pm$ 0.033	&	0.384 $\pm$ 0.023 &	0.381 $\pm$ 0.021	&	0.364 $\pm$ 0.037 \\

LMS2	&	LMS & 20	& 0.452 $\pm$ 0.011	&	0.400 $\pm$ 0.018 &	0.402 $\pm$ 0.011	&	0.391 $\pm$ 0.015 \\

LDA1		&	LDA & 20	& 0.483 $\pm$ 0.034	&	0.379 $\pm$ 0.022 &	0.377 $\pm$ 0.020	&	0.355 $\pm$ 0.038 \\

LDA2	&	LDA & 20	& 0.453 $\pm$ 0.014	&	0.382 $\pm$ 0.015 &	0.387 $\pm$ 0.009	&	0.369 $\pm$ 0.016 \\
\hline
Entropy	&	LMS & 30	& 0.423 $\pm$ 0.030	& 	0.394 $\pm$ 0.028	&	0.394 $\pm$ 0.020 	&	0.385 $\pm$ 0.027	\\

LMS1	&	LMS & 30	& 0.450 $\pm$ 0.018	&	0.395 $\pm$ 0.028	&	0.393 $\pm$ 0.027	&	0.379 $\pm$ 0.050	\\

LMS2	&	LMS & 30	& 0.446 $\pm$ 0.013	&	0.409 $\pm$ 0.020	&	0.408 $\pm$ 0.013	&	0.403 $\pm$ 0.019	\\

LDA1		&	LDA & 30	& 0.430 $\pm$ 0.041	&	0.384 $\pm$ 0.027	&	0.382 $\pm$ 0.027	&	0.353 $\pm$ 0.053	\\

LDA2	&	LDA & 30	& 0.433 $\pm$ 0.021	&	0.397 $\pm$ 0.017	&	0.396 $\pm$ 0.016	&	0.386 $\pm$ 0.026	\\
\hline
\end{tabular}}
\caption{F1-macro (i.e. F1-macro = $\frac{1}{C}\sum_{k\in C}F1_k$, with $C$ as the number of classes for the 9-fold experimental protocol) results, based only on technical features, for the five sorting lists classified based on LMS,LDA, and RBFN for the next $T=$$10^{th}$, $20^{th}$, and $30^{th}$ events, respectively, as the predicted horizon. The number of best features used in the above methods is different in every case (as seen in \hyperref[fig:FeatureNumber]{Fig. 3}). }
\end{table}

\begin{table}[H]\label{tab:Table7}
\centering
\scalebox{0.75}{
\begin{tabular}{ccccccc}
\hline
Sorting  & Classifier & T &$Accuracy$ & $Precision$ & $Recall$ & $F1$  \\
\hline
Entropy	&	LMS & 10	& 0.393 $\pm$ 0.109	& 	0.399 $\pm$ 0.047	&	0.419 $\pm$ 0.047	&	0.340 $\pm$ 0.064	\\

LMS1 	&	LMS & 10	&  0.665 $\pm$ 0.033	&	0.468 $\pm$ 0.043	&	0.388 $\pm$ 0.016	&	0.366 $\pm$ 0.016	\\

LMS2	&	LMS & 10	& 0.571 $\pm$ 0.071		&	0.470 $\pm$ 0.053	&	0.418 $\pm$ 0.032	&	0.384 $\pm$ 0.020	\\

LDA1		& 	LDA & 10	& 0.611 $\pm$ 0.088	&	0.422 $\pm$ 0.039	&	0.390 $\pm$ 0.020	&	0.370 $\pm$ 0.024	\\

LDA2	&	LDA & 10	& 0.380 $\pm$ 0.101	&	0.401 $\pm$ 0.024	&	0.428 $\pm$ 0.027	&	0.339 $\pm$ 0.063	\\
\hline
Entropy	&	LMS & 20	& 0.400 $\pm$ 0.074	& 	0.408 $\pm$ 0.048	&	0.422 $\pm$ 0.048	&	0.372 $\pm$ 0.061	\\

LMS1	&	LMS & 20	& 0.553 $\pm$ 0.029	&	0.429 $\pm$ 0.037	&	0.373 $\pm$ 0.016	&	0.335 $\pm$ 0.025	\\

LMS2	&	LMS & 20	& 0.483 $\pm$ 0.017	&	0.447 $\pm$ 0.022	&	0.457 $\pm$ 0.025	&	0.435 $\pm$ 0.029 \\

LDA1		&	LDA & 20	& 0.513 $\pm$ 0.072	&	0.402 $\pm$ 0.032	&	0.379 $\pm$ 0.026	&	0.347 $\pm$ 0.040	\\

LDA2	&	LDA & 20	& 0.424 $\pm$ 0.073	&	0.431 $\pm$ 0.026	&	0.444 $\pm$ 0.023	&	0.340 $\pm$ 0.053	\\
\hline
Entropy	&	LMS & 30	& 0.410 $\pm$ 0.062	& 	0.416 $\pm$ 0.051	&	0.423 $\pm$ 0.048	&	0.391 $\pm$ 0.060	\\

LMS1	&	LMS & 30	& 0.478 $\pm$ 0.022	&	0.407 $\pm$ 0.038	&	0.370 $\pm$ 0.019	&	0.320 $\pm$ 0.036	\\

LMS2	&	LMS & 30	& 0.481 $\pm$ 0.012	&	0.457 $\pm$ 0.026	&	0.460 $\pm$ 0.024	&	0.449 $\pm$ 0.034	\\

LDA1		&	LDA & 30	& 0.464 $\pm$ 0.037	&	0.394 $\pm$ 0.030	&	0.378 $\pm$ 0.027	&	0.338 $\pm$ 0.049	\\

LDA2	&	LDA & 30	& 0.425 $\pm$ 0.063	&	0.437 $\pm$ 0.029	&	0.443 $\pm$ 0.022	&	0.406 $\pm$ 0.055	\\
\hline
\end{tabular}}
\caption{F1-macro (i.e. F1-macro = $\frac{1}{C}\sum_{k\in C}F1_k$, with $C$ as the number of classes for the 9-fold experimental protocol) results, based only on quantitative features, for the five sorting lists classified based on LMS,LDA, and RBFN for the next $T=$$10^{th}$, $20^{th}$, and $30^{th}$ events, respectively, as the predicted horizon. The number of best features used in the above methods is different in every case (as seen in \hyperref[fig:FeatureNumber]{Fig. 3}). }
\end{table}

\begin{table}[H]\label{tab:Table8}
\centering
\scalebox{0.94}{
\begin{tabular}{cccccccc}
\hline
Sorting &  Classifier &5 & 50 & 100 & 200 & 273  \\
\hline
Entropy 	 & LMS	& 0.319 $\pm$ 0.008 & 0.363 $\pm$ 0.027 & 0.414 $\pm$ 0.020 &  0.425 $\pm$ 0.025 & 0.440 $\pm$ 0.018 \\

LMS1	 & LMS	& 0.377 $\pm$ 0.008 & 0.374 $\pm$ 0.036 & 0.393 $\pm$ 0.047 &  0.419 $\pm$ 0.036 & 0.440 $\pm$ 0.018 \\

LMS2 	& LMS 	& 0.402 $\pm$ 0.015  & 0.443 $\pm$ 0.014 & 0.441 $\pm$ 0.018 &  0.440 $\pm$ 0.018 & 0.440 $\pm$ 0.018 \\

LDA1	 	& LMS 	& 0.373 $\pm$ 0.013  & 0.380 $\pm$ 0.017 & 0.395 $\pm$ 0.016 & 0.315 $\pm$ 0.018 & 0.289 $\pm$ 0.025 \\

LDA2 	& LMS 	& 0.412 $\pm$ 0.011 & 0.420 $\pm$ 0.017 & 0.420 $\pm$ 0.019 & 0.289 $\pm$ 0.013 & 0.309 $\pm$ 0.027 \\

LDA1 	& LMS 	& 0.370 $\pm$ 0.011 & 0.372 $\pm$ 0.032 & 0.387 $\pm$ 0.041 & 0.440 $\pm$ 0.017 & 0.440 $\pm$ 0.018 \\

LDA2 	& LMS 	& 0.421 $\pm$ 0.010 & 0.435 $\pm$ 0.011 & 0.435 $\pm$ 0.014 & 0.440 $\pm$ 0.017 & 0.441 $\pm$ 0.018 \\

Entropy 	& RBFN 	& 0.316 $\pm$ 0.010 & 0.363 $\pm$ 0.020 & 0.413 $\pm$ 0.016 & 0.430 $\pm$ 0.016 & 0.441 $\pm$ 0.016 \\

LMS1 	& RBFN	& 0.387 $\pm$ 0.018 & 0.402 $\pm$ 0.022 & 0.421 $\pm$ 0.017 & 0.429 $\pm$ 0.017 & 0.441 $\pm$ 0.016 \\

LMS2 	& RBFN 	& 0.403 $\pm$ 0.015 & 0.444 $\pm$ 0.012 & 0.439 $\pm$ 0.013 & 0.439 $\pm$ 0.019 & 0.442 $\pm$ 0.016 \\

LDA1 	& RBFN 	& 0.371 $\pm$ 0.011 & 0.387 $\pm$ 0.021 & 0.411 $\pm$ 0.013 & 0.440 $\pm$ 0.016 & 0.441 $\pm$ 0.016 \\

LDA2 	& RBFN 	& 0.416 $\pm$ 0.011 & 0.434 $\pm$ 0.015 & 0.436 $\pm$ 0.015 & 0.436 $\pm$ 0.016 & 0.442 $\pm$ 0.016 \\
\hline
\end{tabular}}
\caption{F1 results based on different numbers of best features for the five criteria of the wrapper-based feature selection methods.}
\end{table}

There is a dual interpretation of the suggested feature lists and wrapper method results. Regarding the feature lists, we have five different feature sorting methods starting from entropy, to LMS1 and LMS2 and continue to LDA1 and LDA2. More specifically, results based on the entropy sorting method reveal that the first 20 places are covered by features almost entirely from technical indicators (i.e. 19 out of 20 places and only one from the first basic group), while the first 100 best places are covered by 36 quant features, 48 technical features, and 16 from the first basic group.

For the LMS case, we present two sorting lists where we use two different criteria for the final feature selection. In the LMS1 case, the first best 20 places covered by features are derived mainly from quantitative analysis (11 out of 20), 7 from the first basic group, and only 2 from the technical group. The first place is covered by a very advanced feature based on the logistic regression model for online learning.  For the same method, the first 100 best places covered by 25 quant features, 18 features from technical analysis, and the remaining 57 from the first basic group. In the LMS2 case, the first 20 best places are covered by 7 features from the quant pool, 9 from the technical pool, and only 4 from the first basic group. LMS2 also selects the advanced feature based on the logistic regression model for online learning first.

The last method that we use as the basis for the feature selection process is based on LDA. In a similar fashion, we use two different criteria as a measure for the selection process. In the LDA1 case, the first 20 best places are covered by 10 quant features, 3 technical indicators, and 7 from the first basic group. The first 100 best positions are covered by 19 quant features, 20 technical features, and the remaining 61 places from the first basic group. Again, the first place is covered by the advanced feature based on the logistic regression model for online learning. The last feature selection model, LDA2, selects 6 features from the quant pool, 6 from the technical pool, and 8 from the first basic group. LDA2 selects for the first best 100 places 24 quant features, 27 technical features, and 49 from the first basic group.
To gain better insight into the feature list, we present the names of the best 10 features for each of the 5 sorting methods in \hyperref[tab:slist]{Table 9}.

\begin{table}
\centering
\scalebox{0.66}{
\begin{tabular}{rl}
\hline
\hline \\
\textbf{Feature Sets} & \textbf{Description} \\
\hline \\
Entropy				& \\
\cline{1-1}
\\
				1	&	Autocorrelation\\
				2	& 	Donchian Channels\\
				3	& 	Highest High\\
				4	& 	Center of Gravity Oscillator \\					
			 	5	&	Heikin-Ashi\\
				6	&	Linear Regr. - Regression Coeffic.\\
				7	&	Linear Regr. - Correlation Coeffic.\\
				8	&	T3\\
				9	&	TEMA\\
				10	&	TRIMA\\					
\hline \\
LMS1 				& \\
\cline{1-1}
\\
				1	&	Logistic Regr. - Local Spatial Ratio  \\
				2	& 	Best LOB Level - Bid Side Volume\\
				3	& 	Second Best LOB Level - Ask Volume\\
				4	& 	Price and Volume Derivation\\					
			 	5	&	Best LOB Level - Ask Side\\
				6	&	Linear Regr. - Corr. Coeffic. \\
				7	&	Logistic Regr. - Logistic Coeffic.\\
				8	&	Logistic Regr. - Extended Spatial Ratio\\
				9	&	Autocorrelation for Log Returns \\
				10	&	Partial Autocorrelation\\	
\hline \\
LMS2 				& \\
\cline{1-1}
\\
				1	&	Logistic Regression - Spatial Ratio \\
				2	& 	Cointegration - Boolean Vector\\
				3	& 	Cointegration - Test Statistics\\
				4	& 	Price and Volume Means\\					
			 	5	&	Average Type Intensity\\
				6	&	Average Type Intensity\\
				7	&	Spread \& Mid-Price\\
				8	&	Alligator Jaw\\
				9	&	Directional Index\\
				10	&	Fractals\\		
\hline \\
LDA1 				& \\
\cline{1-1}
\\
				1	&	Logistic Regression - Spatial Ratio\\
				2	& 	Second Best LOB Level - Ask Volume\\
				3	& 	Price \& Volume derivation\\
				4	& 	Spread \& Mid-Price\\					
			 	5	&	Partial Autocorrelation for Log Returns\\
				6	&	Linear Regression Line - Squared Corr. Coeffic.\\
				7	&	Order Book Imbalance\\
				8	&	Linear Regression - Corr. Coeffic.\\
				9	&	Linear Regression - Regr. Coeffic.\\
				10	&	Third Best LOB Level - Ask Volume\\	
\hline \\
LDA2 				& \\
\cline{1-1}
\\
				1	&	Logistic Regression - Probability Estimation\\
				2	& 	Logistic Regression - Spatial Ratio\\
				3	& 	Bollinger Bands\\
				4	& 	Alligator Teeth\\					
			 	5	&	Cointegration - Test Statistics\\
				6	&	Best LOB Level - Bid Side Volume\\
				7	&	Cointegration - p Values\\
				8	&	Price \& Volume Means\\
				9	&	Price \& Volume derivation\\
				10	&	Price differences\\	
\hline
\hline
\end{tabular}}
\caption{List for the first 10 best features for the 5 sorting methods}
\label{tab:slist}
\end{table}

The second interpretation of our findings is the performance of the 12 different classifiers (based on LMS, LDA, and RBFN) that we used to measure, in terms of F1 score, the predictability of the mid-price movement. \hyperref[fig:BarPlot]{Fig.3} provides a quick overview of the F1 score performance in terms of best feature numbers and classifiers. We can divide these twelve models (pairs based on the sorting and classification method) into three groups according to their response in terms of information flow. The first group, where LMS2-LMS, LDA2-LMS, LMS2-RBFN, and LDA2-RBFN belong, reach their plateau very early in the incremental process of adding less informative features. These models where able to reach their maximum (or close to their maximum) F1 score performance with approximately 5 best features, which means that the dimensionality of the input matrix to the classification model is quite small. The second group of models, Entropy-LMS, LMS1-LMS, LDA1-LMS, Entropy-RBFN, LMS1-RBFN, and LDA1-RBFN, had a slower reaction in the process of reaching their best F1 score performance. The last group of models, LDA1-LDA and LDA2-LDA, reached their best performance very early in the process (which is not higher performance compared to the other models) with only five features. Interestingly, it is right after this point that their predictability power starts to decrease.

The experiments conducted show that this quantitative analysis can provide significant trading information, but results are improved with respect to features also taken from the technical pool. Features on the top of the lists come from the logistic regression model. This is the very first time this model is presented as a feature in the HFT sphere. This shows that more sophisticated features from the pool of quantitative analysis will provide the ML trader with vital information regarding metrics prediction. We would like to point out that the main idea of the present work is to utilize the majority of the technical indicators \footnote{In the literature it is very common to see experiments based on a limited number of technical indicators.} and provide a fair evaluation against other state-of-the-art features and advanced quantitative hand-crafted features like the adaptive logistic regression feature and see which ones are more informative. Classification performance can be easily improved by utilizing more advanced classifiers like convolutional neural networks and recurrent neural networks (e.g., \cite{dixon2018sequence}), but it is outside of the scope of our evaluation. Our work open avenues for other application as well. For instance, the same type of analysis is suitable for exchange rates and bitcoin time series. Last, we intend to test our experimental protocol on a longer trading period. 

\section{Conclusion}\label{SS:Con}
\noindent In this paper, we extracted hand-crafted features inspired by technical and quantitative analysis and tested their validity on the mid-price movement prediction task. We used entropy, least-mean-squares (LMS), and linear discriminant analysis (LDA) criteria to guide feature selection methods combined with linear and non-linear classifiers based on LMS, LDA, and RBFN. This work is the first attempt of this extent to develop a framework in information edge discovery via informative hand-crafted features. Therefore, we provided the description of three sets of hand-crafted features adjusted to the HFT universe by considering each 10-message book block as a separate trading unit (i.e. trading days). We evaluated our theoretical framework on five ITCH feed data stocks from the Nordic stock market. The dataset contained more than 4.5 million events and was incorporated into the hand-crafted features. The results suggest that sorting methods and classifiers can be combined in such a way that market makers and traders can reach, with only very few informative features, the best performance of their algorithm. Furthermore, the very advanced quantitative feature based on logistic regression for online learning is placed first among most the five sorting methods. This is a strong indication for future research on developing more advanced features combined with more sophisticated feature selection methods.

%

\appendix

\medskip
\medskip
\medskip
\medskip
\noindent\textbf{Appendix}

\renewcommand{\thesection}{\Alph{section}}

\section{Feature Pool}\label{SS:Feat}
\renewcommand{\thesection}{\Alph{section}}

\subsection{First Group of Features}
This set of features is based on \cite{doi:10.1080/14697688.2015.1032546} and \cite{DBLP:journals/corr/NtakarisMKGI17} and is divided into three groups: basic, time-insensitive, and time-sensitive. These are fundamental features since they reflect the raw data directly without any statistical analysis or interpolation. We calculated them as follows:
\subsubsection{Basic}
\begin{itemize}
\item $u_1$ = $\{ P_i^{ask}, V_i^{ask}, P_i^{bid}, V_i^{bid}\}_{i=1}^n$											\end{itemize}
which represents the raw data of the 10 levels of our LOB.

\subsubsection{Time-Insensitive}
\begin{itemize}
\item $u_2$ = $\{(P_i^{ask}-P_i^{bid}), (P_i^{ask}+P_i^{bid})/2 \}_{i=1}^n$ \\
\item  $u_3 = \\
 \{P_n^{ask}-P_1^{ask}, P_1^{bid}-P_n^{bid}, |P_{i+1}^{ask}-P_i^{ask}|, |P_{i+1}^{bid}-P_i^{bid}| \}_{i+1}^n$\\
\item $u_4$ = $\Big\{ \frac{1}{n}\sum\limits_{i=1}^{n}P_i^{ask},  \frac{1}{n}\sum\limits_{i=1}^{n}P_i^{bid},  \frac{1}{n}\sum\limits_{i=1}^{n}V_i^{ask},  \frac{1}{n}\sum\limits_{i=1}^{n}V_i^{bid}\Big\}$	\\
\item $u_5 = \Big\{ \sum\limits_{i=1}^{n}(P_i^{ask} - P_i^{bid}), \sum\limits_{i=1}^{n}(V_i^{ask} - V_i^{bid})  \Big\}$		\end{itemize}
where $u_3$ represents the spread and the mid-price, $u_4$ the price differences, and $u_5$ the price and the volume means, respectively.

\subsubsection{Time-Sensitive}			
\begin{itemize}	
\item $u_6$ = $\Big\{dP_i^{ask}/{dt}, dP_i^{bid}/{dt}, dV_i^{ask}/{dt}, dV_i^{bid}/{dt} \Big\}_{i=1}^n $		
\item $u_7$ = $\Big\{ \lambda_{\Delta t}^{1}, \lambda_{\Delta t}^{2}, \lambda_{\Delta t}^{3}, \lambda_{\Delta t}^{4}, \lambda_{\Delta t}^{5}, \lambda_{\Delta t}^{6}  \Big\}$
\item $u_8 = \Big\{ \textbf{1}_{\lambda_{\Delta_t}^{1}>\lambda_{\Delta_T}^{1}}, \textbf{1}_{\lambda_{\Delta_t}^{2}>\lambda_{\Delta_T}^{2}}, \textbf{1}_{\lambda_{\Delta_t}^{3}>\lambda_{\Delta_T}^{3}}, \textbf{1}_{\lambda_{\Delta_t}^{4}>\lambda_{\Delta_T}^{4}},\\ \textbf{1}_{\lambda_{\Delta_t}^{5}>\lambda_{\Delta_T}^{5}}, \textbf{1}_{\lambda_{\Delta_t}^{6}>\lambda_{\Delta_T}^{6}} \Big\}$ \\
\item $u_9$ = $\{d\lambda^{1}/dt, d\lambda^{2}/dt, d\lambda^{3}/dt, d\lambda^{4}/dt, d\lambda^{5}/dt, \\d\lambda^{6}/dt \}$
\end{itemize}
where $u_6$ represents the price and volume derivation, $u_7$ the average type intensity, $u_8$ the relative comparison intensity, and $u_9$ the limit activity acceleration, respectively.

\subsection{Technical Analysis}\label{SS:Tech}

\noindent Technical analysis is based mainly on the idea that historical data provides all the relevant information for trading prediction. The prediction, based on technical analysis, takes place according to open-close-high and low prices in day-to-day trading. We adjust this idea to the HFT ML problem for every 10-MB block of events. More specifically, we consider every 10-MB block as a 'trading' day (i.e. with \textit{t} as the current 10-MB block and \textit{t}-1 the previous 10-MB block), and we extract features according to this formation as follows:
\bigbreak

\subsubsection{Accumulation Distribution Line}
\bigbreak
\noindent Accumulation Distribution Line (ADL) \citep{chua2006sammy} is a volume-based indicator for measuring supply and demand and is a three-step process: \\

\begin{itemize}
\item $Money Flow Multiplier$ = [$(C_{t} -  L_{t}) - (H_{t} - C_{t})] /(H_{t} - L_{t})$
\item $Money Flow Volume_{t}$ = $Money Flow Multiplier$ x $Block Period Volume$
\item $ADL$ = $ADL_{t-1} + Money Flow Volume_{t}$
\end{itemize}
\noindent with $C_{t}, L_{t}$, and $H_{t}$ being the closing, lowest, and highest current 10-MB block prices, respectively, and $Block Period Volume$, $ADL_{t-1}$, and $Money Flow Volume_{t}$ are the total amounts of 10-MB block volume density, the previous ADL price, and the current $Money Flow Volume_{t}$, respectively.
\bigbreak

\subsubsection{Awesome Oscillator}
\bigbreak

\noindent An awesome oscillator (AO) \citep{williams1998new} is used to capture market momentum. Here, we adjust the trading rules according to the previous block horizon investigation to 5 and 34 previous 10-MB blocks as follows:\\

\begin{itemize}
\item AO = $SMA_{5}((H_{t} + L_{t})/2) - SMA_{34}((H_{t} + L_{t})/2)$\\
\end{itemize}
\noindent where $SMA_{5}$ and $SMA_{34}$ are the simple moving averages of the previous 5 and 34 previous blocks, respectively.

\bigbreak
\subsubsection{Accelerator Oscillator}
\bigbreak

\noindent An accelerator oscillator \citep{williams1998new} is another market momentum indicator derived from AO. It is calculated as follows:\\

\begin{itemize}
\item $AC = AO - SMA_{5}(AO)$
\end{itemize}

\bigbreak
\subsubsection{Average Directional Index}
\bigbreak

\noindent An average directional index (ADX) indicator \citep{wilder1986relative} has been developed to identify the strength of a current trend. The ADX is calculated as follows:

\begin{itemize}
\item $TR = max(H_{t} - L_{t}, |H_{t} - CL_{t-1}|, |L_{t} - CL_{t-1}| )$
\item $+DM = H_{t} - H_{t-1} $
\item $-DM = L_{t} - L_{t-1}$
\item $TR_{14} =TR_{t-1} - (TR_{t-1}/14) + TR$
\item $+DM_{14} = (+DL_{t-14}) - ((+DL_{t-14})/14) + (+DM)$
\item $-DM_{14} = (-DL_{t-14}) - ((-DL_{t-14})/14) + (-DM)$
\item $ +DI_{14} =  100 \times ((+D_{14})/(+TR_{14}))$
\item $ -DI_{14} =  100 \times ((-D_{14})/(-TR_{14}))$
\item $ DI_{diff_{14}} = |(+D_{14}) - (-D_{14})|$
\item $ DI_{sum_{14}}= |(+D_{14}) + (-D_{14})|$
\item $ DX = 100 \times ((DI_{diff_{14}})/(DI_{sum_{14}}))$
\item $ADX = (ADX_{t-1} \times 13) + DX)/14$
\end{itemize}

\noindent
where TR = true range, $H_{t}$ = the current 10-block's highest MB price, $L_{t}$ = the current 10-block's lowest MB price, $CL_{t}$ = the previous 10-block's closing MB price, $+DM =$ positive Directional Movement (DM), $-DM$ = negative DM, $ TR_{14}$ =TR based on the previous 14-blocks, $TR_{t-1}$ = the previous TR price, $+DM_{14}$ =  DM based on the previous 14 $+DM$ blocks, $-DM_{14}$ =  DM based on the previous 14 $-DM$ blocks, $+DM_{t-14}$ =  +DM of the previous 14 $+DM$ blocks, $DI_{diff_{14}}$ = is the directional indicator (DI) of the difference between $+DM_{14}$ and $-DM_{14}$, $DI_{sum_{14}}$ = DI of the sum between $+DM_{14}$ and $-DM_{14}$, $DX$ = directional movement index and $ADX_{t-1}$ = the previous average directional index.

\bigbreak
\subsubsection{Average Directional Movement Index Rating }
\bigbreak

\noindent An average directional movement index rating (ADXR) evaluates the momentum change of ADX, and it is calculated as the average of the current and previous price of ADX:

\begin{itemize}
\item $ADXR = (ADX + ADX_{t-1})/2$
\end{itemize}

\bigbreak
\subsubsection{Displaced Moving Average Based on Williams Alligator Indicator}
\bigbreak

\noindent A displaced moving average \citep{gregory2012trading} is the basis for building a trading signal named Alligator. In practice, this is a combination of three moving averages (MA). We adjust this idea as follows:

\begin{itemize}
\item $Alligator_{Jaw}$ = $SMA_{13}((H_t + L_t)/2)$
\item $Alligator_{Teeth}$ = $SMA_{8}((H_t + L_t)/2)$
\item $Alligator_{Lips}$ = $SMA_{5}((H_t + L_t)/2)$
\end{itemize}
where $SMA_{13}((H_t + L_t)/2)$, $SMA_{8}((H_t + L_t)/2)$, and $SMA_{5}((H_t + L_t)/2)$ are the simple moving averages based on the previous 13, 8, and 5 average highest and lowest block prices, respectively.

\bigbreak
\subsubsection{Absolute Price Oscillator }
\bigbreak

\noindent An absolute price oscillator (APO) belongs to the family of price oscillators. It is a comparison between fast and slow exponential moving averages and is calculated as follows:

\begin{itemize}
\item $M_t$ = $(H_t + L_t)/2$
\item APO = $EMA_{5}(M_t) - EMA_{13}(M_t)$
\end{itemize}
where $EMA_{5}(M_t)$ and $EMA_{13}(M_t)$ are the exponential moving averages of range 5 and 13 periods, respectively, for the average of high and low prices of the current 10-MB block.

\bigbreak
\subsubsection{Aroon Indicator}
\bigbreak

\noindent An Aroon indicator \citep{chande1994new} is used as a measure of trend identification of an underlying asset. More specifically, the indicator has two main bodies: the uptrend and downtrend calculation. We calculate the Aroon indicator based on the previous twenty 10-MB blocks for the highest-high and lowest-low prices, respectively, as follows:

\begin{itemize}
\item $Arron_{Up}$ = (20 - $H_{high_{20}}/20) \times 100$
\item $Arron_{Down}$ = (20 - $L_{low_{20}}/20) \times 100$
\end{itemize}
where $H_{high_{20}}$ and $L_{low_{20}}$ are the highest-high and lowest-low 20 previous 10-MB block prices, respectively.

\bigbreak
\subsubsection{Aroon Oscillator}
\bigbreak

\noindent An Aroon oscillator is the difference between $Aroon_{Up}$ and $Aroon_{Down}$ indicators, which makes their comparison easier:

\begin{itemize}
\item Arron Oscillator = $Aroon_{Up}$ - $Aroon_{Down}$
\end{itemize}

\bigbreak
\subsubsection{Average True Range }
\bigbreak

\noindent Average true range (ATR) \citep{wilder1978new} is a technical indicator which measures the degree of variability in the market and is calculated as follows:

\begin{itemize}
\item ATR = $(ATR_{t-1} \times (N-1) + TR)/N$
\end{itemize}
Here we use N =14, where N is the number of the previous 10-TR values, and $ATR_{t-1}$ is the previous ATR 10-MB block price.

\bigbreak
\subsubsection{Bollinger Bands}
\bigbreak

\noindent Bollinger bands \citep{bollinger2001bollinger} are volatility bands which focus on the price edges of the created envelope (middle, upper, and lower band) and can be calculated as follows:

\begin{itemize}
\item $BB_{middle}$ = $SMA_{20}(CL)$
\item $BB_{upper}$ = $SMA_{20}(CL) + BB_{std_{20}} \times 2$
\item $BB_{lower}$ = $SMA_{20}(CL) - BB_{std_{20}} \times 2$
\end{itemize}
where $BB_{middle}, BB_{upper}$, and $BB_{lower}$ represent the middle, upper, and lower Bollinger bands, $SMA_{20}(CL)$ represents the simple moving average of the previous twenty 10-block closing prices, and $BB_{std_{20}}$ represents the standard deviation of the last twenty 10-MB blocks.

\bigbreak
\subsubsection{Ichimoku Clouds}
\bigbreak

\noindent Ichimoku clouds \citep{muranaka2000ichimoku} are 'one glance equilibrium charts,' which means that the trader can easily identify a good trading signal and is possible since this type of indicator contains dense information (i.e. momentum and trend direction). Five modules are used in an indicator's calculation:

\begin{itemize}
\item Conversion Line ($Tenkan-sen$) = $(H_{9} + L_{9})/2$
\item Base Line ($Kijun-sen$) = $H_{26} + L_{26}$
\item Leading Span A ($Senkou$ $Span$ $A$) = (Conversion Line + Base line)/2
\item Leading Span B ($Senkou$ $Span$ $B$) = $(H_{52} + L_{52})/2$
\item Lagging Span ($Chikou$ $Span$) = $CL_{26}$
\end{itemize}
\noindent where H, L, and CL denotes the highest, lowest, and closing prices of the 10-MB raw data where subscripts 9, 26, and 52 denotes the historical horizon of our trading rules.

\bigbreak
\subsubsection{Chande Momentum Oscillator }
\bigbreak

\noindent A Chande momentum oscillator (CMO) \citep{chande1994new} belongs to the family of technical momentum oscillators and can monitor overbought and oversold situations. There are two modules in the calculation process:

\begin{itemize}
\item $S_{u}$ = $\sum\limits_{i = 1}^{19}CL_{i} \times \mathds{1}_{CL_{t}>CL_{t-19}}$
\item $S_{d}$ = $\sum\limits_{i = 1}^{19}CL_{i} \times \mathds{1}_{CL_{t}<CL_{t-19}}$
\item CMO = 100 $\times$ $(S_{u}$ - $S_{d})/(S_{u}$ + $S_{d}) $
\end{itemize}
where $CL_{i}$ is the 10-block's closing price with $i = 1$, and $CL_{t}$ and $CL_{t-19}$ are the current block's closing price and the 19 previous blocks� closing prices, respectively.

\bigbreak
\subsubsection{Chaikin Oscillator}
\bigbreak

\noindent The main purpose of a Chaikin oscillator \citep{naiman2009small} is to measure the momentum of the 	accumulation distribution line as follows:

\begin{itemize}
\item $MFM $= $(CL_{t} - L_{t}) - (H_{t} - CL_{t})]/(H_{t}-L_{t})$
\item $MFV$ = $MFM \times \sum\limits_{j = 1}^{10}V_{j}$
\item $ADL$ =$ ADL_{t-1} + MFM$
\item Chaikin Oscillator = $EMA_3(ADL)$ - $EMA_{10}(ADL)$
\end{itemize}
where $MFM$ and $MFV$ stand for \textit{Money Flow Multiplier} and \textit{Money Flow Volume}, respectively, V is the volume of each of the trading events in the 10-block MB, and $EMA_3(ADL)$ and $EMA_{10}(ADL)$ are the exponential moving average for the past 3 and 10 10-MB blocks, respectively.

\bigbreak
\subsubsection{Chandelier Exit}
\bigbreak
\noindent A Chandelier exit \citep{elder2002come} is part of the trailing stop strategies based on the volatility measured by the ATR indicator. It is separated based on the number of ATRs that are below the 22-period high (long) or above the 22-period low (short) and is calculated as follows:

\begin{itemize}
\item $Chandelier_{Long}$ = $H_{22} - ATR_{22} \times 3$
\item $Chandelier_{Short}$ = $L_{22} + ATR_{22} \times 3$
\end{itemize}
where $H_{22}$ and $L_{22}$ denote the highest and lowest prices for a period of 22 10-MB blocks, and $ATR_{22}$ are the ATR values for the 22 previous 10-MB blocks.

\bigbreak
\subsubsection{Center of Gravity Oscillator }
\bigbreak

\noindent A center of gravity oscillator (COG) \citep{ehlers2001rocket} is a comparison of current prices against older prices within a specific time window and is calculated as follows:

\begin{itemize}
\item $M_t$ = $(H_{t} + L_{t})/2$
\item COG = $- (M_t + r \times M_{t-1})/ (M_t + M_{t-1})$
\end{itemize}
where $M_{t}$ is the current mid-price of the highest and lowest prices of each of the 10-MB blocks, and r is a weight that increases according to the number of the previous $M_{t-1}$ prices.

\bigbreak
\subsubsection{Donchian Channels }
\bigbreak

\noindent A Donchian channel (DC) \citep{rayome2007technical} is an indicator which bands the signal and notifies the ML trader of a price breakout. There are three modules in the calculation process:

\begin{itemize}
\item $DC_{upper}$ = $H_{high_{20}}$
\item $DC_{lower}$ = $L_{low_{20}}$
\item $DC_{middle}$ = $(H_{high_{20}} + L_{low_{20}})/2$
\end{itemize}
where $H_{high_{20}}$ and $L_{low_{20}}$ are the highest high and lowest low prices of the previous twenty 10-MB blocks.

\bigbreak
\subsubsection{Double Exponential Moving Average  }
\bigbreak

\noindent A double exponential moving average (DEMA) \citep{mulloy1994smoothing} provides a smoothed average and offers a diminished amount of delays as follows:

\begin{itemize}
\item $M_t$ = $(H_{t} + L_{t})/2$
\item $DEMA$ = $2 \times EMA_{20}(M_t) - EMA_{20}(EMA_{20}(M_t))$
\end{itemize}
where $EMA_{20}$ is the exponential moving average of span 20 of the closing prices under the 10-MB block format.

\bigbreak
\subsubsection{Detrended Price Oscillator }
\bigbreak

\noindent A detrended price oscillator (DPO) is an indicator used for short-term and long-term signal identification. A DPO eliminates cycles which are longer than the MA horizon. On day-to-day trading, the closing prices are considered for the calculation, but here, we use the highest 10-MB block price as follows:

\begin{itemize}
\item $DPO$ = $(H_{high_{10}}/(10+2)) - SMA_{10}(CL)$.
\end{itemize}

\bigbreak
\subsubsection{Heikin-Ashi}
\bigbreak

\noindent Heikin-Ashi \citep{valcu2004using} is a candlestick method and is described as a visual technique that eliminates irregularities:

\begin{itemize}
\item $Heikin_{Close}$ = $(O_{t} + H_{t} + L_{t} + CL_{t})/4$
\item $Heikin_{Open}$ = $(O_{t-1} + CL_{t-1})/2$
\item $Heikin_{High}$ = $ max(H_{t}, O_{t-1}, CL_{t-1})$
\item $Heikin_{Low}$ = $min(L_{t}, O_{t-1}, CL_{t-1})$
\end{itemize}
where $O_{t-1}$ and $CL_{t-1}$ are the open and close prices of the previous 10-MB block.

\bigbreak
\subsubsection{Highest High and Lowest Low}
\bigbreak

\noindent Highest high and lowest low creates an envelope of the trading signal for the last twenty 10-MB blocks:

\begin{itemize}
\item $Highest_{High}$ = $H_{high_{20}}$
\item $Lowest_{Low}$ = $L_{low_{20}}$
\end{itemize}

\bigbreak
\subsubsection{Hull MA}
\bigbreak

\noindent A Hull moving average is a weighted moving average that reduces the smoothing lag effect by using the square root of the block period. It is calculated as follows:

\begin{itemize}
\item $Hull_{MA}$ = $WMA_{\sqrt{10}}(AHL)(2 \times WMA_5(AHL) - WMA_{10}(AHL))$
\end{itemize}
where $WMA_{5}(AHL)$ and $WMA_{10}(AHL)$ denote the weighted moving average of the average high and low 10-MB block for periods 5 and 10, respectively.

\bigbreak
\subsubsection{Internal Bar Strength }
\bigbreak

\noindent Internal bar strength (IBS) \citep{pagonidisibs} is based on the position of the day’s closing price in relation to the day’s range where we adjust this idea to the 10-MB block setup as follows:

\begin{itemize}
\item $IBS$ = $(CL_{t} - L_{t})/(H_{t}-L_{t})$.
\end{itemize}

\bigbreak
\subsubsection{Keltner Channels}
\bigbreak

\noindent Keltner channels \citep{keltner1960make} are based on Bollinger bands. The main difference, for this volatility-based indicator, is that it uses ATR instead of standard deviation, as follows:

\begin{itemize}
\item $Middle_{Channel}$ = $EMA_{{20_{AHL}}}$
\item $Upper_{Channel}$ = $Middle_{Channel} + 2 \times ATR_{10}$
\item $Lower_{Channel}$ = $Middle_{Channel} - 2 \times ATR_{10}$.
\end{itemize}

\bigbreak
\subsubsection{Moving Average Convergence/Divergence Oscillator (MACD)}
\bigbreak

\noindent A moving average convergence/divergence oscillator \citep{aspray1989individual} is a measure of the convergence and divergence of two moving averages and is calculated as follows:

\begin{itemize}
\item $MACD$ = $EMA_{12}(AHL) - EMA_{26}(AHL)$
\end{itemize}
where AHL is the average of high and low prices for 12 and 26 previous 10-MB blocks, respectively, with $EMA_{12}(AHL)$ and $EMA_{26}(AHL)$ as the exponential moving average of AHL of span 12 and 26, respectively.

\bigbreak
\subsubsection{Median Price}
\bigbreak

\noindent Median price is an indicator which simplifies the price overview. We calculate this indicator based on the 10-MB block highest and lowest average prices:

\begin{itemize}
\item $Median_t = (H_{t} + L_{t})/2$
\end{itemize}

\bigbreak
\subsubsection{Momentum }
\bigbreak

\noindent A momentum (MOM) indicator measures the rate of change of the selected time series. In our case, we calculate it based on closing prices:

\begin{itemize}
\item $MOM$ = $CL_{t} - CL_{t-1}$.
\end{itemize}

\bigbreak
\subsubsection{Variable Moving Average }
\bigbreak

\noindent A variable moving average (VMA) \citep{chande1992adapting} is a dynamic indicator which acts as a variable-length moving average with volatility-adaptation capabilities. We calculate VMA based on the efficiency ratio (ER) as follows:

\begin{itemize}
\item $Direction$ = $|CL_{t} - CL_{t-3}|$
\item $Volatility$ = $3 \times \sum\limits_{ii = 1}^{3} |CL_{ii} - CL_{ii+1}|$
\item $ER$ = $Direction/Volatility$
\item $VMA$ = $\sum\limits_{jj=1}^{3}\alpha \times ER_{jj} \times CL_{jj}$
\end{itemize}
where $\alpha = 2/(N+1)$, for N = 3 previous 10-MB blocks, is the adaptive parameter.

\bigbreak
\subsubsection{Normalized Average True Range }
\bigbreak

\noindent A normalized average true range (NATR) normalizes the average true range as follows:

\begin{itemize}
\item $NATR$ = $(ATR / CL_{t}) \times 100$.
\end{itemize}

\bigbreak
\subsubsection{Percentage Price Oscillator }
\bigbreak

\noindent A percentage price oscillator (PPO) displays the convergence and divergence of two moving averages and focuses on the percentage change of the larger moving average, as follows:

\begin{itemize}
\item $MACD$ = $EMA_{12}(AHL) - EMA_{26}(AHL)$
\item $PPO$ = $(MACD/EMA_{26}(AHL)) \times 100$.
\end{itemize}

\bigbreak
\subsubsection{Rate of Change }
\bigbreak

\noindent Rate of change (ROC) measures the ascent or descent speed of the time series change:

\begin{itemize}
\item $ROC$ = 	$(CL_{t} - CL_{t-12}/CL_{t-12}) \times 100$.
\end{itemize}

\bigbreak
\subsubsection{Relative Strength Index }
\bigbreak

\noindent A relative strength index (RSI) \citep{wilder1986relative} is a measure of the velocity and magnitude of directional time series movements and is calculated as follows:

\begin{itemize}
\item $CL_{d}$ = $CL_{t} - CL_{t-1}$
\item $AG_{14}$ = $\sum\limits_{l = 1}^{14} CL_{d_{l}} \mathds{1}_{CL_{d_{t}}>CL_{d_{t-1}}}$
\item $AL_{14}$ = $\sum\limits_{l = 1}^{14} CL_{d_{l}} \mathds{1}_{CL_{d_{t}}<CL_{d_{t-1}}}$
\item $Relative_{Strength}$ = $AG_{14}/ AL_{14}$
\item $RSI$ = $100-100/(1+Relative_{Strength})$
\end{itemize}
where $AG_{14}$ and $AL_{14}$ denotes the average gain and loss of the last fourteen 10-MB blocks, respectively.

\bigbreak
\subsubsection{Parabolic Stop and Reverse }
\bigbreak

\noindent Parabolic SAR (PSAR) \citep{wilder1978new} is a trend following indicator which protects profits. There are two main modules for its calculation, the Rising SAR and the Falling SAR, and they are calculated as follows:

\begin{itemize}
\item Rising SAR
\begin{itemize}
\item $AF$ = $Incremental$  $increase$ $of$ $a$ $predefined$ $step$
\item $EP$ = $H_{High_{5}}$
\item $SAR$ = $SAR_{t-1} + AF_{t-1}(EP_{t-1} - SAR_{t-1})$
\end{itemize}
\item Falling SAR
\begin{itemize}
\item $AF$ = $Incremental$  $increase$ $of$ $a$ $predefined$ $step$
\item $EP$ = $L_{Low_{5}}$
\item $SAR$ = $SAR_{t-1} - AF_{t-1}(EP_{t-1} - SAR_{t-1})$
\end{itemize}
\end{itemize}
where $AF$ is the acceleration factor, and $EP$ is the extreme point

\bigbreak
\subsubsection{Standard Deviation}
\bigbreak

\noindent Standard deviation is a measure of volatility. We calculate this indicator based on the closing prices of every 10-MB block, as follows:

\begin{itemize}
\item $Deviation$ = $CL_{t} - SMA_{10}(CL)$
\item $SASD$ = $\sqrt{SMA_{10}(SVD)}$
\end{itemize}
where $SMA_{10}(CL)$ is the simple moving average of the last 10 closing 10-MB prices, $SASD$ is the squared deviation of the SMA of the standard deviation (SVD) of the last 10 closing values of our 10-MB blocks.

\bigbreak
\subsubsection{Stochastic Relative Strength Index }
\bigbreak

\noindent A stochastic relative strength index (Stoch RSI) \citep{chande1994new} is a range-bound momentum oscillator which provides information for the RSI based on the closing prices in terms of high and low stock prices:

\begin{itemize}
\item $Stoch_{RSI}$ = $(RSI_{curr} - RSI_{L_{Low_{10}}})/(RSI_{H_{High_{10}}} - RSI_{L_{Low_{10}}})$
\end{itemize}
where $RSI_{L_{Low_{10}}}$ and $RSI_{H_{High_{10}}}$ are the lowest low and highest high of the last ten RSI values.

\bigbreak
\subsubsection{T3-Triple Exponential Moving Average }
\bigbreak

\noindent A triple exponential moving average \citep{tillson1998better} is a moving average indicator where the main motivation for its development is to reduce lag in the time series response. For this reason, we use the closing prices for our calculation and perform a reversal explanation calculation as follows:

\begin{itemize}
\item $T3$ = $c_{1} \times EMA_{6} + c_{2} \times EMA_{5} + c_{3} \times EMA_{4} + c_{4} \times EMA_{3}$
\end{itemize}
with:
\begin{itemize}
\item $c_{1}$ = $-\alpha^{3}$
\item $c_{2}$ = $3\times \alpha^{2} + 3\times \alpha^{3}$
\item $c_{3}$ = $– 6\times \alpha^{2} – 3 \times \alpha – 3 \times \alpha^{3}$
\item $c_{4}$ = $1 + 3\times \alpha + \alpha^{3} + 3\times \alpha^{2}$
\item $EMA_{1}$ = $EMA_{10}(CL)$
\item $EMA_{2}$ = $EMA_{10}(EMA_{1})$
\item $EMA_{3}$ = $EMA_{10}(EMA_{2})$
\item $EMA_{4}$ = $EMA_{10}(EMA_{3})$
\item $EMA_{5}$ = $EMA_{10}(EMA_{4})$
\item $EMA_{6}$ = $EMA_{10}(EMA_{5})$
\end{itemize}
where $\alpha$ is the volume factor, and $EMA_{10}(CL)$ is the exponential moving average of the 10 previous 10-MB closing prices.

\bigbreak
\subsubsection{Triple Exponential Moving Average }
\bigbreak

\noindent A triple exponential moving average (TEMA) \citep{mulloy1994smoothing} is an attempt to reduce the lag associated with MA by adding weight to the most recent prices:

\begin{itemize}
\item $TEMA$ = $(3 \times EMA_{10}(CL)) - (3 \times EMA_{10}(EMA_{10}(CL)) + EMA_{10}(EMA_{10}(EMA_{10}(CL)))$
\end{itemize}
with EMA being, in every case, the exponential moving average of the previous 10 prices (i.e. previous EMA and closing prices).

\bigbreak
\subsubsection{Triangular Moving Average }
\bigbreak

\noindent A triangular moving average (TRIMA) is the average of the time series with emphasis placed on the middle region:

\begin{itemize}
\item $TRIMA$ = $SMA_{10}(SMA_{10}(SMA_{10}(CL)))$
\end{itemize}
Where, for its calculation, we use the closing prices of the last 10 10-MB blocks.
\bigbreak
\subsubsection{Triple Exponential Average }
\bigbreak

\noindent A triple exponential average (TRIX) is a momentum oscillator which measures the rate of change of the triple smoothed moving average as follows:

\begin{itemize}
\item $EMA_{First}$ = $EMA_{10}(CL)$
\item $EMA_{Double}$ = $EMA_{10}(EMA_{First})$
\item $EMA_{Triple}$ = $EMA_{10}(EMA_{Double})$
\item $TRIX$ = 1-$period$ $Rate$ $of$ $Change$.
\end{itemize}

\bigbreak
\subsubsection{True Strength Index }
\bigbreak

\noindent A true strength index (TSI) \citep{blau1991double} is an indicator which specifies the overbought and oversold levels with market return anticipation. We calculate TSI as follows:

\begin{itemize}
\item $PC$ = $CL_{k} - CL_{k-1}$, where $k  = 2,..., T$
\item $APC$ = $|CL_{k} - CL_{k-1}|$, where $k  = 2,..., T$
\item $EMA_{1}$ = $EMA_{25}(PC)$
\item $EMA_2$ = $EMA_{13}(EMA_1)$
\item $EMA_{3}$ = $EMA_{25}(APC)$
\item $EMA_4$ = $EMA_{13}(EMA_3)$
\item $TSI$ = $100 \times EMA_{2}/ EMA_{4}$
\end{itemize}
where PC represents the closing price differences for the whole time series lookback period.

\bigbreak
\subsubsection{Ultimate Oscillator }
\bigbreak

\noindent An ultimate oscillator (UO) \citep{williams1985ultimate} is a momentum oscillator indicator with a multiple timeframe perspective. There are three main modules as presented in the following calculations:

\begin{itemize}
\item Average of seven 10-MB blocks
\begin{itemize}
\item $BP$ = $CL_{t} - (CL_{t-1}\mathds{1}_{CL_{t-1}<L_{t}} + L_{t}\mathds{1}_{CL_{t-1}>L_{t}})$
\item $TR_1$ =  $CL_{t-1}\mathds{1}_{CL_{t-1}>H_{t}} + H_{curr}\mathds{1}_{CL_{t-1}<H_{t}}$
\item $TR_2$ = $CL_{t-1}\mathds{1}_{CL_{t-1}<L_{t}} + L_{t}\mathds{1}_{CL_{t-1}>L_{t}}$
\item $TR$ = $TR_1 + TR_2$
\item $Average_7$ = $\sum\limits_{l= 1}^{7}BP_l / \sum\limits_{l= 1}^{7}TR_l$
\end{itemize}

\item Average of fourteen 10-MB blocks
\begin{itemize}
\item $BP$ = $CL_{t} - (CL_{t-1}\mathds{1}_{CL_{t-1}<L_{t}} + L_{t}\mathds{1}_{CL_{t-1}>L_{t}})$
\item $TR_3$ =  $CL_{t-1}\mathds{1}_{CL_{t-1}>H_{t}} + H_{t}\mathds{1}_{CL_{t-1}<H_{t}}$
\item $TR_4$ = $CL_{t-1}\mathds{1}_{CL_{t-1}<L_{t}} + L_{t}\mathds{1}_{CL_{t-1}>L_{t}}$
\item $TR$ = $TR_3 + TR_4$
\item $Average_{14}$ = $\sum\limits_{l= 1}^{14}BP _l/ \sum\limits_{l= 1}^{14}TR_l$
\end{itemize}

\item Average of twenty-eight 10-MB blocks
\begin{itemize}
\item $BP$ = $CL_{t} - (CL_{t-1}\mathds{1}_{CL_{t-1}<L_{t}} + L_{t}\mathds{1}_{CL_{t-1}>L_{t}})$
\item $TR_5$ =  $CL_{t-1}\mathds{1}_{CL_{t-1}>H_{t}} + H_{t}\mathds{1}_{CL_{t-1}<H_{t}}$
\item $TR_6$ = $CL_{t-1}\mathds{1}_{CL_{t-1}<L_{t}} + L_{t}\mathds{1}_{CL_{t-1}>L_{t}}$
\item $TR$ = $TR_5 + TR_6$
\item $Average_{28}$ = $\sum\limits_{l= 1}^{28}BP_l / \sum\limits_{l= 1}^{28}TR_l$
\end{itemize}

\item $UO$ = $100 \times \big[ (4 \times Average_7) + (2 \times Average_{14}) + Average_{28}\big] / (4+2+1)$
\end{itemize}
where $BP$ represents buying pressure.

\bigbreak
\subsubsection{Weighted Close }
\bigbreak

\noindent Weighted close (WCL) is the average of the four universal types of prices which are included in each of our 10-MB blocks:

\begin{itemize}
\item $WCL$ = $(H_{t} + L_{t}+2 \times CL_{t})/4$.
\end{itemize}

\bigbreak
\subsubsection{Williams \%R}
\bigbreak

\noindent Williams \%R \citep{williams2011long} is a momentum technical indicator which informs the ML trader whether the market is trading close to the high or low trading range. It is calculated as follows:

\begin{itemize}
\item $\%R$ = $-100 \times (H_{High_{14}} - CL_{t})/ (H_{High_{14}} - L_{Low_{14}})$
\end{itemize}
where -100 corrects the inversion.

\bigbreak
\subsubsection{Zero-Lag Exponential Moving Average }
\bigbreak

\noindent Zero-lag exponential moving average (ZLEMA) belongs to the EMA family of indicators where the main purpose is to reduce or remove the impulse lag by introducing an error term. It is calculated as follows:

\begin{itemize}
\item $error$ = $CL - CL_{lag}$
\item $Input$ = $CL + error$
\item $ZLEMA$ = $EMA_{10}(Input)$
\end{itemize}
where $lag = (N-1)/2$ with N = 1 in our case.

\bigbreak
\subsubsection{Fractals}
\bigbreak

\noindent A fractal \citep{gregory2012trading} is an indicator used to detect top and bottom trends by focusing on five consecutive blocks, which, in our case, are five 10-MB blocks used for two different scenarios:

\begin{itemize}
\item $Buy$ $Fractals$

A buy fractal is a sequence of five consecutive 10-MB blocks where the highest high is preceded by two lower highs and is followed by two lower highs.

\item $Sell$ $Fractals$

The opposite framework is a sell fractal. 10-MB blocks can overlap in the quest of these two types of fractals.
\end{itemize}
Here, we calculate fractals separately for the open, close, lowest, and highest 10-MB block prices.

\bigbreak
\subsubsection{Linear Regression Line}
\bigbreak

\noindent Linear regression line (LRL) is a basic statistical method that provides information for a future projection wherein trading is used to capture overextended price trends. We perform LRL for each 10-MB block without any prior stationarity assumptions. The basic calculations are as follows:

\begin{itemize}
\item $PV$ = $c_1 + c_2 \times MB_{prices}$
\item $c_2$ = $r \times (std_{PV}/std_{MB_{prices}})$
\item $r$ = $\frac{\Big(\sum\limits_{i=1}^{10}(MB_{prices}(i) - \overline{MB_{prices}})(PV(i) - \overline{PV})\Big)}{\Big(\sqrt{\sum\limits_{i=1}^{10}(MB_{prices}(i) - \overline{MB_{prices}})^{2} (\sum\limits_{i=1}^{10}(PV(i) - \overline{PV})^{2} \Big)}}$
\item $c_1$ = $\overline{PV} - 	c_2 \times \overline{MB_{prices}}$
\end{itemize}
where $PV$ are the predicted values, r is the correlation coefficient, and $\overline{MB_{prices}}$ and $\overline{PV}$ are the mean of 10-MB block prices and predicted values, respectively.

\bigbreak
\subsubsection{Digital Filtering: Rational Transfer Function}
\bigbreak

\noindent A rational transfer function \citep{62245} is a representation of a linear time-invariant (LTI) filter, with the assumption that the input signal depends on the time-frequency domain, which describes the input-output relationship of a signal. In the Z-tranform domain, we have the following rational transfer function:

\begin{itemize}
\item $O(z)$ = $\frac{b(1) + b(2)z^{-1} + ... +b(n_{b} + 1) + z^{-n_{b}}}{1 + \alpha(2)z^{-1} + ... + \alpha(n_{a} +1)z^{-n_{\alpha}}}I(z)$,
\end{itemize}
where:
\begin{itemize}
\item $I(z)$ and $O(z)$ are the input (i.e. 10-MB block closing prices) and output respectively,
\item $b$ are the numerator coefficients,
\item $\alpha$ are the denominator coefficients,
\item $n_a$ is the feedback order,
\item $n_b$ is the feedforward order,
\item $z$ is the complex variable,
\item the lookback period for the calculations is ten 10-MB blocks.
\end{itemize}

\bigbreak
\subsubsection{Digital Filtering: Savitzky-Golay Filter }
\bigbreak

\noindent A Savitzky-Golay (S-G) digital filter \citep{savitzky1964smoothing}, \cite{5888646} is a discrete convolution with a specific impulse response. We describe how the ML trader can obtain the S-G signal based on higher degree polynomials:

\begin{itemize}
\item Least-Square Filter
\begin{itemize}
\item Objective: minimize error $\mathcal{E}_{N} = \sum\limits_{i = l}^{m}w_i(y_i - \sum\limits_{r = 0}^{n}p_rx_{i}^{r})^{2} $
\item Partial derivative of the polynomial coefficients: \\
$\frac{\partial Q}{\partial p_k} = 0 \Rightarrow \sum\limits_{i=l}^{m}w_i\sum\limits_{r = 0}^{n}p_rx_i^{r+k} = \sum\limits_{i = l}^{m}w_iy_ix_i^{k}$
\item Finite time series allow order summation change: \\
 $\sum\limits_{r = 0}^{n}p_r\sum\limits_{i = l}^{m}w_ix_i^{r+k} = \sum\limits_{i = l}^{m}w_iy_ix_i^{k}$
\item As a result, the desired linear equations are the following:\\

$\begin{bmatrix}
   \sum\limits_{i=l}^{m}w_ix_i^{0} & \sum\limits_{i=l}^{m}w_ix_i^{1} & \dots  & \sum\limits_{i=l}^{m}w_ix_i^{n}  \\
   \sum\limits_{i=l}^{m}w_ix_i^{1} & \sum\limits_{i=l}^{m}w_ix_i^{2} & \dots  & \sum\limits_{i=l}^{m}w_ix_i^{n+1} \\
   \vdots & \vdots  & \ddots & \vdots \\
   \sum\limits_{i=l}^{m}w_ix_i^{n} & \sum\limits_{i=l}^{m}w_ix_i^{n+1} & \dots  & \sum\limits_{i=l}^{m}w_ix_i^{2n}
\end{bmatrix}$ $\begin{bmatrix}
   p_0  \\
   p_1 \\
   \vdots \\
   p_n
\end{bmatrix}$ = \\
\\

$\begin{bmatrix}
   \sum\limits_{i=l}^{m}w_iy_ix_i^{0}  \\
   \sum\limits_{i=l}^{m}w_iy_ix_i^{1}  \\
   \vdots  \\
   \sum\limits_{i=l}^{m}w_iy_ix_i^{n}
\end{bmatrix}$ \\
\\
equivalent to the notation $\textbf{A}\textbf{P} = \textbf{B}$ where matrix $\mathbf{A}^{-1} \in \mathbb{R}^{(n+1) \times(n+1)}$ under the condition that the polynomial degree is $n \leqslant m-l$.
\end{itemize}

\item S-G Filter
\begin{itemize}
\item Local convolution coefficients calculation \\
\\

$\begin{bmatrix}
   p_0  \\
   p_1 \\
   \vdots \\
   p_n
\end{bmatrix} = \\ \begin{bmatrix}
   \sum\limits_{i=l}^{m}w_ix_i^{0} &  \dots  & \sum\limits_{i=l}^{m}w_ix_i^{n}  \\
   \sum\limits_{i=l}^{m}w_ix_i^{1} &  \dots  & \sum\limits_{i=l}^{m}w_ix_i^{n+1} \\
   \vdots & \ddots & \vdots \\
   \sum\limits_{i=l}^{m}w_ix_i^{n}  & \dots  & \sum\limits_{i=l}^{m}w_ix_i^{2n}
\end{bmatrix}^{-1} \begin{bmatrix}
   \sum\limits_{i=l}^{m}w_iy_ix_i^{0}  \\
   \sum\limits_{i=l}^{m}w_iy_ix_i^{1}  \\
   \vdots  \\
   \sum\limits_{i=l}^{m}w_iy_ix_i^{n}
\end{bmatrix}$ $\Rightarrow$ \\
\\

$\begin{bmatrix}
   p_0  \\
   p_1 \\
   \vdots \\
   p_n
\end{bmatrix} =  \begin{bmatrix}
   c_{0,0} & c_{0,1}& \dots  & c_{0,n}  \\
   c_{1,0} & c_{1,1}& \dots  & c_{1,n}  \\
   \vdots & \vdots  & \ddots & \vdots \\
   c_{n,0} & c_{n,1}& \dots  & c_{n,n}  \\
\end{bmatrix} \begin{bmatrix}
   \sum\limits_{i=l}^{m}w_iy_ix_i^{0}  \\
   \sum\limits_{i=l}^{m}w_iy_ix_i^{1}  \\
   \vdots  \\
   \sum\limits_{i=l}^{m}w_iy_ix_i^{n}
\end{bmatrix}$
\\
\item Response at the local point of 0 degree is:\\
\\
$y[0]$ = $c_{0,0} \ast \sum\limits_{i = l}^{m}w_iy_ix_i^{0} + c_{0,1} \ast \sum\limits_{i = l}^{m}w_iy_ix_i^{1} + ... +c_{0,n} \ast \sum\limits_{i = l}^{m}w_iy_ix_i^{n}$\\
\end{itemize}
\end{itemize}

\bigbreak
\subsubsection{Digital Filtering: Zero-Phase Filter }
\bigbreak

\noindent A zero-phase filter \citep{steven1999scientist} is a bidirectional filtering technique. With zero phase slope and even impulse response $h(n)$, the filter provides an output signal, which is a zero phase recursive signal. This method is suitable for our experimental protocol since we use training and testing sets rather than online learning architecture as we will do in \hyperref[sssec:On]{4.2.4}. The calculation process is as follows:

\begin{itemize}
\item Real Impulse response: $h(n)$, $n \in \mathbb{Z}$
\item Discrete-time Fourier Transformation: \\
\\
$H_{\omega T}(h) = \sum\limits_{n = 1}^{\infty}h(n) cos(\omega n T)$ - $j\sum\limits_{n = 1}^{\infty}h(n) sin(\omega n T)$
\item Based on Euler formula and $h$-even: $H(e^{j\omega T}) = \sum\limits_{n = 1}^{\infty}h(n) cos(\omega n T)$
\end{itemize}

\bigbreak
\subsubsection{Remove Offset and Detrend}
\bigbreak

\noindent We present three detrend methods for short-term cycle isolation and calculate them as follows:

\begin{itemize}
\item Remove Offset
\begin{itemize}
\item $Offset$ = $CL_{t} - (\sum\limits_{l = 1}^{n}CL_l)/n$\\
where $n$ denotes the 10-MB lookback period
\end{itemize}
\item Detrend - Least Squares Fitting Line
\begin{itemize}
\item $\textbf{R}^2$ =  $\sum\limits_{i =1}^{n }[y_i - g(x_i)]^2$
\item $\frac{\partial (\textbf{R}^2)}{\partial \alpha} = 0$
\item $\frac{\partial (\textbf{R}^2)}{\partial b} = 0$
\item
$\begin{bmatrix}
   \alpha \\
   b \\
\end{bmatrix} = \begin{bmatrix}
   n & \sum\limits_{i=l}^{n}x_i  \\
   \sum\limits_{i=l}^{n}x_i  & \sum\limits_{i=l}^{n}x_i^2   \\
\end{bmatrix}^{-1} \begin{bmatrix}
    \sum\limits_{i=l}^{n}y_i   \\
   \sum\limits_{i=l}^{n}x_iy_i   \\
\end{bmatrix}$\\
\\
where $\alpha$ and $b$ are the regression coefficients of $g$, and $x$ represents the 10-MB closing prices.
\end{itemize}
\end{itemize}

\bigbreak
\subsubsection{Beta-like Calculation}
\bigbreak

\noindent Beta \citep{french2003treynor} is a volatility indicator which considers market risk. We adjust the notion of beta calculation to our experimental protocol where we index based on the average of the closing prices ($Av_{CL}$) with $Av_{t}$ as the current MB block price and $Av_{t-1}$ as the previous MB block's closing price. Our calculations are as follow:

\begin{itemize}
\item $Index_{CL}$ = $CL_{t}/CL_{t-1}$
\item $Index_{Av_{CL}}$ = $Av_{t}/Av_{t-1}$
\item $Dev_{CL}$ = $Index_{CL} - SMA_{10}(Index_{CL})$
\item $Dev_{Av_{CL}}$ = $Index_{Av_{CL}} - SMA_{10}(Index_{Av_{CL}})$
\item $Beta$ = $cov_{10}(Dev_{CL}, Dev_{Av_{CL}})/ var_{10}(Dev_{Av_{CL}})$
\end{itemize}
where $cov_{10}(Dev_{CL}, Dev_{Av_{CL}})$ represents the covariance between the current closing price and the average of the previous ten 10-MB closing prices, and $var_{10}(Dev_{Av_{CL}})$ is the variance of the sum of the ten previous $Index_{Av_{CL}}$.

\subsection{Quantitative Analysis}\label{SS:Quant}

\noindent Quantitative analysis captures trading activity mainly via statistical modelling. We focus on time series analysis, and more specifically, we examine features such as autocorrelation and partial autocorrelation, among others (e.g., statistical tests), while in the end of the section, we build an ML feature extraction method based on an online learning setup and test the validity of our hypothesis.

\bigbreak
\subsubsection{Autocorrelation and Partial Correlation}
\bigbreak

\noindent Autocorrelation and partial correlation \citep{box2015time}, \citep{eshel2003yule} are key features in the development of time series analysis. We treat our time series (i.e. stock prices and log returns per 10-MB blocks) as stationary stochastic processes since we estimate their local behavior based on 10-MB blocks:

\begin{itemize}
\item Autocorrelation
\begin{itemize}
\item $ac_{k}$ = $\frac{\textit{E}[(z_t - \mu)(z_{t+k} - \mu)]}{\sqrt{\textit{E}[(z_t - \mu)^2]\textit{E}[(z_{t+k} - \mu)^2]}}$
\end{itemize}
where $z_t$ and $z_{t+k}$ are the time series of lag $k$, $\mu = E[z_t] = \int_{-\infty}^{\infty}zp(z)dz$ and $\sigma_{z}^2 = \textit{E}[(z_t - \mu)^2] = \int_{-\infty}^{\infty}(z - \mu)^{2}p(z)dz$ are the constant mean and constant variance respectively.
\item Partial Correlation
\begin{itemize}
\item For the general case of an autoregressive model $AR(p)$, we have:\\
$x_{i+1}$ = $\phi_1x_i + \phi_2x_{i-1} +...+ \phi_px_{i-p+1} + \xi_{i+1}$ of lag 1 up to $p$ follows:\\
\begin{itemize}
\item $<x_{i}x_{i+1}>$ = $\sum\limits_{j = 1}^{p}(\phi_j<x_ix_{i-j+1}>)$
\item $<x_{i-1}x_{i+1}>$ = $\sum\limits_{j = 1}^{p}(\phi_j<x_{i-1}x_{i-j+1}>)$
\item $<x_{i-k+1}x_{i+1}>$ = $\sum\limits_{j = 1}^{p}(\phi_j<x_{i-k+1}x_{i-j+1}>)$
\item $<x_{i-p+1}x_{i+1}>$ = $\sum\limits_{j = 1}^{p}(\phi_j<x_{i-p+1}x_{i-j+1}>)$

by diving with $N-1$ and autocovariance of zero separated periods (where the autocovariance function is even), all the lag periods above will be:\\
\item $r_1$ = $\sum\limits_{j=1}^{p}\phi_jr_{j-1}$
\item $r_2$ = $\sum\limits_{j=1}^{p}\phi_jr_{j-2}$
\item $r_k$ = $\sum\limits_{j=1}^{p}\phi_jr_{j-k}$
\item $r_p$ = $\sum\limits_{j=1}^{p}\phi_jr_{j-p}$
 \end{itemize}
where $2.1.5$, $2.1.6$, $2.1.7$ and $2.1.8$ can be described by the matrix operations $\pmb{R}\pmb{\Phi} = \pmb{r}$, $\pmb{R} \in \mathbb{R}^{p \times p}$, $\pmb{\Phi} \in \mathbb{R}^{p \times 1}$ and $\pmb{r} \in \mathbb{R}^{p \times 1}$. The symmetric and full rank $\pmb{\Phi}$ are as follows:
$\hat{\pmb{\Phi}}$ = $ \pmb{R}^{-1}\pmb{r}$.

\item \textit{Yule-Walker} Equations calculation:\\

\begin{itemize}
\item Lag interval $1\leqslant i \leqslant p$
\item $\hat{\pmb{\Phi}}$ = $\Big(\pmb{R}^{(i)} \Big)^{-1}\pmb{r}^{(i)} = \begin{bmatrix}
   \hat{\phi_1} \\
    \hat{\phi_2} \\
   \vdots \\
    \hat{\phi_i}
\end{bmatrix}$
\end{itemize}

\end{itemize}
\end{itemize}

\bigbreak
\subsubsection{Cointegration}
\bigbreak

\noindent We investigate time-series equilibrium \citep{hamilton1994time}, \citep{10.2307/1913236} by testing the cointegrated hypothesis. Utilizing the cointegration test will help ML traders avoid the problem of spurious regression. We employ the Engle-Granger (EG) test for the multivariable case of LOB ask ($A_t$) and bid ($B_t$) times series. We formulate the EG test for the ask and bid LOB prices as follows:

\begin{itemize}
\item $A_t$ and $B_t$ $\mathtt{\sim}$  $I(d)$, where $I(d)$ represents the order of integration
\item Cointegration equation based on the error term: $u_t = A_t  - \alpha B_t$
\item EG Hypothesis: $u(t) \mathtt{\sim} I(d), d\neq 0$
\item Perform ordinary leat squares (OLS) for the estimation of $\hat{\alpha}$ and unit root test for: $\hat{u} = A_t - \hat{\alpha}B_t$
\end{itemize}

\bigbreak
\subsubsection{Order Book Imbalance}
\bigbreak

\noindent We calculate the order book imbalance \citep{sirignano2016deep} based on the volume depth of our LOB as follows:

\begin{itemize}
\item $VI$ = $\frac{V_l^b - V_l^{\alpha}}{V_l^b +  V_l^{\alpha}}$
\end{itemize}
where $V_l^{\alpha}$ and $V_l^{b}$ are the volume sizes for the ask and bid LOB sides at level $l$.

\section{Feature Sorting Lists}\label{SS:App}
A detailed feature name list is available upon request.

\begin{enumerate}
\item Features sorting list based on Entropy: \\
\\
\{217;157;165;154;164;207;209;190;191;192;193;174;182;183;146;161;171;172;173;156;155;\\
184;194;137;177;176;138;195;136;218;213;181;147;245;243;247;188;241;242;244;240;246;\\
255;248;249;189;210;265;211;226;236;225;235;221;231;223;233;222;232;214;169;220;\\
230;228;238;224;234;83;84;227;237;139;135;134;142;162;140;185;129;86;128;186;212;\\
141;250;261;16;20;262;153;14;81;12;208;260;4;18;10;259;24;196;2;163;150;187;82;28;\\
197;22;8;26;6;32;30;36;34;40;148;175;160;149;38;151;60;59;58;57;56;55;54;53;52;51;\\
198;215;216;158;167;159;168;152;166;100;102;21;25;29;13;17;9;5;33;1;23;19;15;27;11;31;\\
37;7;3;35;85;39;104;252;96;98;106;269;108;92;112;110;88;253;116;124;120;145;94;50;\\
90;180;256;114;49;170;118;123;122;48;126;268;119;115;80;47;111;143;144;46;107;70;45;\\
125;103;121;44;117;43;113;99;42;109;254;270;271;105;41;95;101;91;272;97;273;93;179;\\
69;178;87;68;79;89;127;67;78;201;199;66;133;77;65;76;205;203;61;200;202;71;75;62;\\
72;64;206;204;257;63;132;131;74;73;130;251;258;267;266;219;229;239;263;264\}
\\
\item Features sorting list based on LMS1:\\
\\
\{269;4;6;88;2;211;266;267;249;250;251;92;253;254;91;252;255;193;122;270;174;183;\\
268;108;103;32;18;147;216;100;118;263;264;111;41;90;112;20;273;24;127;116;120;109;\\
110;126;225;235;164;107;98;102;124;165;89;94;133;114;119;104;96;95;87;115;188;61;\\
93;125;101;97;105;113;208;209;99;180;121;117;246;106;123;189;248;228;238;8;210;\\
186;130;185;10;14;242;157;136;16;218;240;244;65;66;64;69;153;28;73;22;170;143;\\
142;184;178;30;154;76;79;67;75;63;74;78;256;247;245;146;219;229;239;68;187;62;\\
70;176;201;179;194;72;197;131;132;77;217;42;43;44;45;46;47;48;49;50;71;85;207;40;\\
258;226;236;80;260;262;134;135;36;204;144;145;38;26;12;84;199;195;182;215;156;\\
171;158;151;167;148;161;168;191;152;159;160;149;150;141;198;169;166;1;212;213;181;\\
3;5;7;9;11;13;15;17;19;21;23;25;27;29;31;33;35;37;39;51;52;53;54;55;56;57;58;59;60;\\
81;82;86;155;163;196;175;214;272;172;173;140;190;192;139;200;162;227;237;222;232;\\
220;230;241;243;224;234;271;206;34;83;177;205;203;221;223;231;233;202;138;137;\\
261;128;129;257;259;265\}
\\
\item Features sorting list based on LMS2:\\
\\
\{269;259;262;83;129;130;49;137;177;205;203;257;223;202;200;199;243;273;176;206;\\
256;204;265;132;10;2;14;84;170;78;240;226;182;157;61;80;242;217;41;70;50;207;165;\\
150;164;93;87;62;43;89;66;215;18;154;251;111;222;8;261;201;258;270;271;65;96;151;\\
216;272;210;186;124;120;153;94;187;92;211;117;109;101;162;166;29;213;184;185;198;\\
195;127;146;191;192;193;196;174;171;159;149;161;139;125;113;106;102;266;118;104;\\
218;36;38;156;190;250;63;85;133;12;121;90;34;40;175;91;248;241;227;245;152;128;\\
189;178;214;136;142;158;224;225;112;99;115;264;212;169;141;163;220;221;188;197;\\
194;209;208;168;22;105;114;110;268;16;23;181;140;119;123;100;126;122;260;244;\\
246;134;135;131;56;103;173;167;6;228;47;97;255;107;180;71;155;4;254;253;179;\\
82;138;32;28;143;252;116;30;144;147;88;108;73;95;98;249;20;51;160;247;55;59;\\
5;148;42;7;76;31;54;3;145;77;46;19;231;48;17;15;81;232;21;45;52;230;236;37;1;24;\\
58;69;13;53;35;67;172;33;183;79;86;26;267;75;219;25;234;9;44;39;11;229;237;57;\\
235;239;60;27;68;64;74;233;238;72;263\}
\\
\item Features sorting list based on LDA1:\\
\\
\{269;6;88;41;255;211;266;250;249;10;252;251;253;268;8;108;114;174;193;254;100;\\
263;264;110;186;273;216;90;99;122;185;92;183;267;16;225;235;14;103;119;112;107;\\
95;104;147;111;91;115;270;127;109;116;120;18;89;94;118;126;98;180;106;208;209;124;\\
96;188;113;125;121;153;123;105;117;93;97;101;248;242;61;133;189;87;102;210;145;66;\\
65;64;69;136;184;142;73;76;157;75;74;78;67;63;79;170;178;77;219;229;239;262;182;\\
130;245;70;22;194;244;24;12;265;84;247;167;173;146;60;207;59;17;33;196;158;165;\\
4;218;25;149;203;3;36;53;37;86;21;30;155;58;164;48;246;161;223;26;85;226;205;43;\\
144;80;47;15;135;179;152;27;160;39;38;81;241;50;236;40;220;7;204;260;83;143;258;\\
168;166;51;141;162;23;57;19;131;9;42;132;82;56;49;62;154;128;5;228;259;55;181;191;\\
163;156;187;272;213;224;52;46;35;1;54;234;169;150;240;227;45;238;31;201;192;190;\\
199;261;172;44;134;2;140;129;20;72;214;215;195;68;151;271;198;237;171;11;29;137;221;\\
222;32;13;217;148;230;232;231;233;197;28;159;206;139;212;256;176;177;243;71;200;\\
34;138;175;202;257\}
\\
\item Features sorting list based on LDA2:\\
\\
\{265;269;257;138;259;4;262;83;108;68;129;205;204;120;6;48;248;141;179;203;212;\\
139;184;43;144;118;79;177;18;52;8;193;132;110;70;191;100;103;146;241;143;206;252;\\
247;273;63;211;207;22;10;142;244;16;258;122;221;219;87;217;47;93;140;40;180;202;\\
34;256;2;115;96;218;134;102;99;270;111;253;66;189;88;90;94;36;199;12;75;254;243;\\
72;137;45;272;64;251;77;222;155;255;210;104;209;174;267;105;194;50;14;126;109;32;\\
170;200;125;98;127;89;227;44;201;119;28;245;61;65;268;192;216;112;20;186;42;250;\\
187;107;121;116;84;185;128;30;237;156;124;160;195;133;147;41;223;215;123;113;135;\\
173;148;271;214;169;131;232;39;149;35;178;71;190;31;198;106;157;188;38;260;168;\\
153;228;55;5;69;246;114;67;15;266;76;152;33;183;37;27;238;46;242;17;166;101;54;\\
23;117;58;56;11;167;261;9;91;162;29;7;97;163;151;233;57;78;24;95;86;225;164;220;\\
154;181;249;171;230;229;130;172;60;26;182;51;1;3;136;159;25;59;208;145;85;53;80;\\
224;92;240;13;81;231;175;264;197;150;74;158;234;213;196;176;235;19;21;263;165;82;\\
226;236;73;161;239;62;49\}
\end{enumerate}

\section*{Acknowledgment}

\noindent The research leading to these results has received funding from the H2020 Project BigDataFinance MSCA-ITN-ETN 675044 (http://bigdatafinance.eu), Training for Big Data in Financial Research and Risk Management.\\

\noindent The authors wish to acknowledge CSC-IT Center for Science, Finland, for generous computational resources.\\

\section*{References}

\bibliography{QuantFinanceB.bib}

\end{document}